\documentclass[prl,superscriptaddress,a4paper,twocolumn]{revtex4-1}
\usepackage{geometry}
\usepackage[english]{babel}
\usepackage[latin1]{inputenc}
\usepackage{hyperref}
\usepackage{amsmath, amssymb, amsfonts, mathrsfs}
\usepackage{graphicx}

\graphicspath{{images/}}
\usepackage{xcolor}
\usepackage{exscale}
\usepackage{graphicx}
\usepackage{amsmath}
\usepackage{latexsym}
\usepackage{amsfonts}
\usepackage{amssymb}
\usepackage{bbm}

\def\pmx{\begin{pmatrix}}
\def\emx{\end{pmatrix}}

\newcommand{\ket}[1]{|#1\rangle}
\newcommand{\bra}[1]{ \langle #1 \,  |}

\begin{document} 

\title{Orbital angular momentum of photons and the entanglement of Laguerre-Gaussian modes}

\author{Mario Krenn}
\email{mario.krenn@univie.ac.at}
\affiliation{Vienna Center for Quantum Science \& Technology (VCQ), Faculty of Physics, University of Vienna, Boltzmanngasse 5, 1090 Vienna, Austria.}
\affiliation{Institute for Quantum Optics and Quantum Information (IQOQI), Austrian Academy of Sciences, Boltzmanngasse 3, 1090 Vienna, Austria.}
\author{Mehul Malik}
\affiliation{Vienna Center for Quantum Science \& Technology (VCQ), Faculty of Physics, University of Vienna, Boltzmanngasse 5, 1090 Vienna, Austria.}
\affiliation{Institute for Quantum Optics and Quantum Information (IQOQI), Austrian Academy of Sciences, Boltzmanngasse 3, 1090 Vienna, Austria.}
\author{Manuel Erhard}
\affiliation{Vienna Center for Quantum Science \& Technology (VCQ), Faculty of Physics, University of Vienna, Boltzmanngasse 5, 1090 Vienna, Austria.}
\affiliation{Institute for Quantum Optics and Quantum Information (IQOQI), Austrian Academy of Sciences, Boltzmanngasse 3, 1090 Vienna, Austria.}
\author{Anton Zeilinger}
\email{anton.zeilinger@univie.ac.at}
\affiliation{Vienna Center for Quantum Science \& Technology (VCQ), Faculty of Physics, University of Vienna, Boltzmanngasse 5, 1090 Vienna, Austria.}
\affiliation{Institute for Quantum Optics and Quantum Information (IQOQI), Austrian Academy of Sciences, Boltzmanngasse 3, 1090 Vienna, Austria.}

\begin{abstract}
The identification of orbital angular momentum (OAM) as a fundamental property of a beam of light nearly twenty-five years ago has led to an extensive body of research around this topic. The possibility that single photons can carry OAM has made this degree of freedom an ideal candidate for the investigation of complex quantum phenomena and their applications. Research in this direction has ranged from experiments on complex forms of quantum entanglement to the interaction between light and quantum states of matter. Furthermore, the use of OAM in quantum information has generated a lot of excitement, as it allows for encoding large amounts of information on a single photon. Here we explain the intuition that led to the first quantum experiment with OAM fifteen years ago. We continue by reviewing some key experiments investigating fundamental questions on photonic OAM and the first steps into applying these properties in novel quantum protocols. In the end, we identify several interesting open questions that could form the subject of future investigations with OAM.  \end{abstract}
\maketitle

The orbital angular momentum of light emerges as a consequence of a spatially varying amplitude and phase distribution. Light beams with OAM have a "twisted" or helical phase structure, where the phase winds azimuthally around the optical axis. The intensity distribution of such beams exhibits a characteristic intensity null at the center due to destructive interference. In 1992, Allen et al. \cite{allen1992orbital} showed that a single photon with such a spiral structure carries a well-defined value of orbital-angular-momentum (OAM). While this was a significant development, it remained unclear whether photons can be entangled in their OAM degree of freedom \cite{arlt1999parametric}. Here we explain the physical intuitions that led to the first OAM entanglement experiment \cite{mair2001entanglement}, and then look into some of the developments since then.

\subsection*{Intuitions behind the first OAM quantum experiment}
From earlier experiments it was known that linear momentum is conserved in spontaneous parametric down conversion processes (SPDC). As Laguerre-Gaussian modes can be decomposed in this basis, one would intuitively expect that OAM modes are conserved as well: The measurement of the OAM of a photon is a projection into a superposition of k-states with a certain phase relation. This also projects its partner photon in the same k-state superposition but with inverse phase, as the sum of the phases of the down-converted pair corresponds to the phase of the pump-photon. Obviously, this intuition needed to be verified experimentally.

While it was not known how to measure the OAM of light at the single photon level, there were methods to create beams with OAM. For example, one could start with a Gaussian mode and send it through a hologram with a helical phase structure and thereby create a beam with OAM. The time-reversed scenario works as well: A beam carrying a non-zero OAM incident on a helical phase hologram with opposite OAM can thus be transformed into a Gaussian mode and can then be coupled into a single-mode fiber and measured with a photon detector. In this manner, the combination of a hologram and a single-mode fiber could act as a mode-filter for single photons carrying OAM (see Figure 1a,b).

With this tool in hand, it could now be tested whether OAM is conserved in the SPDC process. By pumping the nonlinear crystal with a Gaussian mode $\ell=0$ it is expected to find the two down-converted photons having opposite OAM values $\ell_1=-\ell_2$. To show this, photon A was projected onto $\ell_1$ and the OAM of the partner photon was measured. As expected, the partner photon always shows opposite OAM values. To show OAM conservation in general, different pump modes were used and the property $\ell_p$ = $\ell_1$ + $\ell_2$ was verified.

\begin{figure*}[ht!]
\includegraphics[width=0.7 \textwidth]{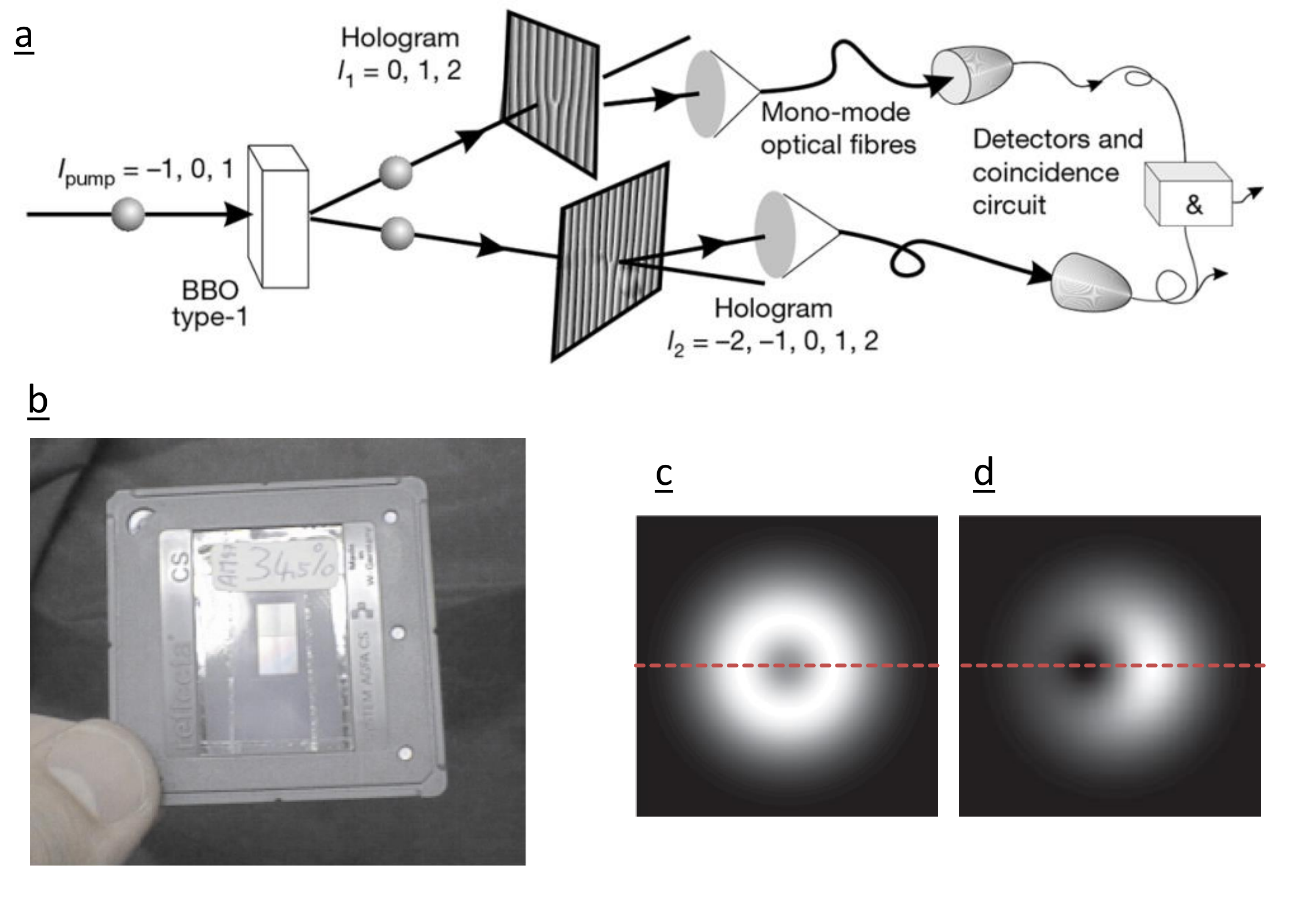}\\
\caption{\textbf{a}: The experimental configuration of Mair et al. \cite{mair2001entanglement} consisted of an SPDC crystal (BBO), which produced photon pairs entangled in OAM. The entanglement was confirmed by using absorption grating holograms which transform the OAM value of the mode. Using a single-mode fiber, only Gauss photons were detected, which allowed the OAM value of single photons to be determined. (Image from \cite{mair2001entanglement}). \textbf{b}: Picture of the absorptive hologram used in the Mair experiment (Image from \cite{mair2000nichtlokale}). \textbf{c} shows the incoherent mixture of a $\ell=0$ and $\ell=1$ mode, while \textbf{c} shows coherent superpositions between those modes (Images from \cite{mair2000nichtlokale}). This property can be used to certify entanglement, by measuring the intensity of one photon along the red depicted line, when its partner photon is projected into a superposition of $\ell=0$ and $\ell=1$.}
\label{fig:fig1Mair}
\end{figure*}

While this showed the conservation of OAM in the SPDC process, it did not demonstrate entanglement yet. To show that the photons did indeed exhibit entanglement in their OAM, one needs to show that they are in a coherent superposition $\ket{\psi}=\frac{1}{\sqrt{2}}\big(\ket{0,0}+\ket{1,1}\big)$, as opposed to an incoherent mixture $\rho=\frac{1}{2}\big(\ket{0,0}\bra{0,0}+\ket{1,1}\bra{1,1}\big)$. The idea was to project one photon into a $\ket{0}+\ket{1}$ superposition and observe the intensity structure of the partner photon. In an incoherent mixture, one would expect some intensity everywhere, because the Gaussian is uniformly non-zero in intensity. However, in stark contrast, a coherent superposition of $\ket{0}+\ket{1}$ will have an intensity null that is shifted from the center to a ring where the amplitude of the $\ket{0}$ and $\ket{1}$ mode are equal to each other (Figure 1c,d). The azimuthal orientation of the null is given by the relative phase between the modes.

The projection into the $\ket{0}+\ket{1}$ superposition was performed with a +1 hologram shifted laterally from the center of the optical axis. (Note that in general, a coherent superposition of $\ket{0}+\ket{\ell}$  has $\ell$  intensity-nulls symmetrically arranged in a ring, thus shifted holograms can only be used to project into superpositions with $\ell=1$). Scanning the transverse intensity profile of the triggered partner photon then showed the intensity null that was expected for a coherent superposition of $\ket{0}$ and $\ket{1}$ modes. In this manner, quantum entanglement of OAM was certified \cite{mair2001entanglement}. 

This work initiated a manifold of experiments investigating single and entangled photons carrying OAM. For example, the distribution of higher-order modes from down-conversion (called spiral bandwidth \cite{torres2003quantum}) is not uniform, which means that higher-order modes have a lower probability of being generated in SPDC than Gaussian modes.  In order to produce maximally entangled high-dimensional states, a filtering technique was developed that increased the amount of entanglement at the cost of reducing the total number of counts \cite{vaziri2003concentration}. Another example of this procedure involved the generation of triggered qutrits for quantum communication \cite{molina2004triggered}. Alternatively, others investigated the replacement of holograms by spiral phase plates \cite{oemrawsingh2005experimental} and sector-plates \cite{oemrawsingh2006high, pors2008shannon, pors2011high} as a simple tool for analyzing LG modes. Spatial mode entanglement was also investigated via Hong-Ou-Mandel interference \cite{walborn2003multimode, peeters2007orbital}.

A significant advance in OAM entanglement experiments was the replacement of static holograms by dynamic spatial light modulators (SLMs) \cite{yao2006observation}. With holograms, only one mode could be measured at a time, rendering many experiments infeasible. SLMs allow the projected mode to be dynamically changed without any realignment of the experiment. This enabled the quick and precise measurement of arbitrary superpositions \cite{jack2010entanglement} of OAM modes, which opened the door to many fundamental experiments with Laguerre-Gauss modes. For example, violations of Bell \cite{leach2009violation}, Leggett \cite{romero2010violation}, and Hardy \cite{chen2012hardy} inequalities were demonstrated with the spatial degree of freedom. Bound states (curious states whose entanglement cannot be distilled) that can only exist in high-dimensional spaces have been created \cite{hiesmayr2013complementarity}. Furthermore, it allowed the investigation of entanglement in even more complex or exotic spatial mode structures carrying OAM, such as Bessel modes \cite{mclaren2012entangled, mclaren2013two}, optical vortex links \cite{romero2011entangled}, and Ince-Gauss modes \cite{krenn2013entangled}. The application of this tool is now standard in entanglement experiments, and was necessary for the realization of many of the experiments discussed below.

\begin{figure*}[ht!]
\includegraphics[width=0.88 \textwidth]{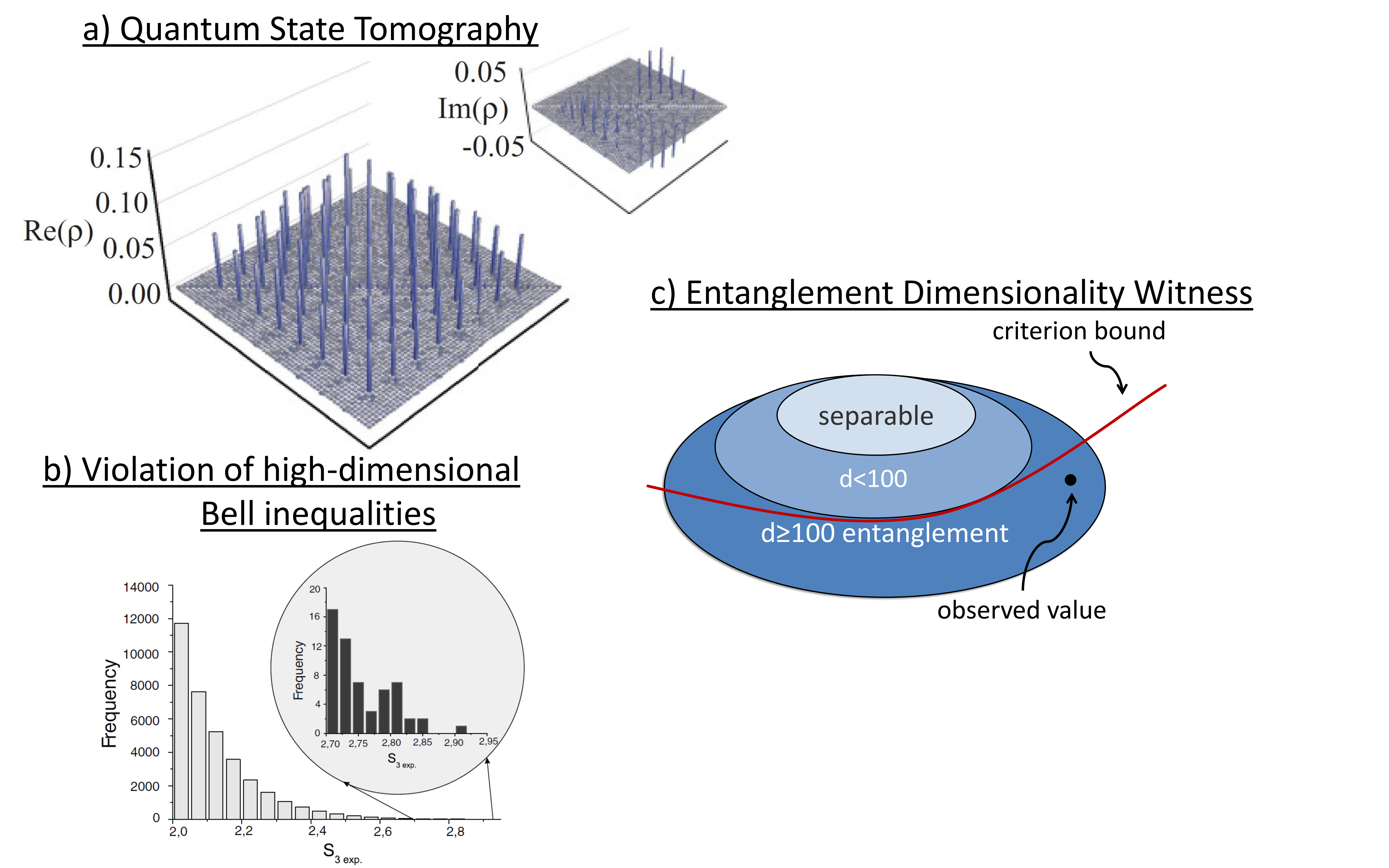}\\
\caption{\underline{Three different methods to investigate high-dimensional entanglement.} \textbf{a}: Quantum State Tomography, \textbf{b}: Violation of high-dimensional Bell inequalities, \textbf{c}: Entanglement Dimensionality Witnesses. \textbf{a}) Quantum tomography, while experimentally and computationally expensive, gives the maximal possible information about the quantum state. The highest-dimensional two-photon state for which quantum tomography (without assumption of state properties) was reported is an 8-dimensionally entangled state. The figure shows the reconstructed density matrix (Image from \cite{agnew2011tomography}). \textbf{b}) Generalized Bell-inequalities can also be used to verify high-dimensional entanglement. These values from \cite{vaziri2002experimental} show the violation of a 3-dimensional Bell-inequality by exceeding the classical bound of $S_3=2$ for various settings in the experiment. \textbf{c}) An entanglement dimensionality (Schmidt Number) witness gives a set of measurements and bounds that a state with entanglement dimension d can maximally reach. If an experiment exceeds this bound, the state was at least (d+1)-dimensionally entangled. In this image, a situation is depicted where a measurement leads to an observed value exceeding the bound for d=100-dimensional entanglement (Image from \cite{krenn2014generation})}  
\label{fig:fig2HighDim}
\end{figure*}

\subsection*{High-dimensional entanglement}
The creation of high-dimensional entanglement is a significant motivation for investigating Laguerre-Gauss modes in the quantum regime. The Mair experiment \cite{mair2001entanglement} discussed above was the first to hint at the possibility of high-dimensional entanglement with OAM. In the years following it, several experiments were performed that quantified OAM-entanglement in more concrete ways. Below, we discuss some of these experiments and the different methods they used for verifying high-dimensional entanglement. 

In order to characterize correlations in higher dimensions and recognize their usefulness for potential applications and fundamental tests, it is important to distinguish between correlations in a high-dimensional space, entanglement in a high-dimensional space and genuine high-dimensional entanglement.

A classically correlated state in three dimensions can be written as 
\begin{align}
\rho_{1d}=\frac{1}{3}\Big(\ket{0,0}\bra{0,0} + \ket{1,1}\bra{1,1} + \ket{2,2}\bra{2,2}\Big).
\label{densityMat3dimSR1}
\end{align}
Such states can be produced by a classical source that produces two photons carrying $\ell=0$ sometimes, and $\ell=1$ or $\ell=2$ at other times, resulting in a completely incoherent mixture. For example, two-dimensional entangled states can be written in the following way:
\begin{align}
\ket{\psi_{2d}^{0,1}}=\frac{\ket{0,0}+\ket{1,1}}{\sqrt{2}}\nonumber\\
\ket{\psi_{2d}^{0,2}}=\frac{\ket{0,0}+\ket{2,2}}{\sqrt{2}}\nonumber\\
\ket{\psi_{2d}^{1,2}}=\frac{\ket{1,1}+\ket{2,2}}{\sqrt{2}}, 
\label{2dimentanglemenet}
\end{align}
An incoherent mixture between these three states,
\begin{align}
\rho_{2d}=\frac{1}{3}\Big(\ket{\psi_{2d}^{0,1}}\bra{\psi_{2d}^{0,1}} + \ket{\psi_{2d}^{0,2}}\bra{\psi_{2d}^{0,2}} + \ket{\psi_{2d}^{1,2}}\bra{\psi_{2d}^{1,2}}\Big),
\label{densityMat3dimSR2}
\end{align}
has correlations in three dimensions, however it is still only two-dimensional entangled, i.e. its Schmidt number is two. The important challenge is to distinguish it from genuine three-dimensional entangled states such as
\begin{align}
\ket{\psi_{3d}}=\frac{\ket{0,0}+\ket{1,1}+\ket{2,2}}{\sqrt{3}}\nonumber\\
\label{3dimentanglemenet}
\end{align}

One method of certifying that a quantum state is high-dimensionally entangled is to reconstruct it via quantum state tomography \cite{langford2004measuring, agnew2011tomography, giovannini2013characterization} and perform a numerical optimisation for the most likely state that yields a physical density matrix (Figure 2a). The resulting matrix can then be analysed with different entanglement measures. The largest two-photon reconstruction performed in this manner was with an 8-dimensionally entangled state \cite{agnew2011tomography}, which required 14,400 measurements over a time of 40 hours. The number of measurements grows rapidly for larger d-dimensional states (the number of measurements scales with $O(d^4)$ for a d-dimensional system) and quickly becomes infeasible, even with optimised versions of tomography \cite{giovannini2013characterization}. Furthermore, the final state optimisation step is computationally expensive and can almost take the same time as the measurements themselves. If prior knowledge is provided, the number of measurements can be vastly decreased. By using compressive sensing techniques, it became possible to reconstruct the quantum state of a 17-dimensional two-photon state \cite{tonolini2014reconstructing}.

The knowledge of the full state is not necessary for extracting information about high-dimensional entanglement (Figure 2b). A different approach is to violate a high-dimensional generalisation of Bell inequalities known as the Collins-Gisin-Linden-Massar-Popescu (CGLMP) inequality \cite{kaszlikowski2000violations, collins2002bell}. This was first demonstrated with fixed holograms in a follow-up of the Mair experiment \cite{vaziri2002experimental}, which for the first time verified that photons from SPDC are entangled in three dimensions of OAM. A much-improved version of this experiment using SLMs and deterministic measurements settings was performed in 2011, and showed the violation of an eleven-dimensional Bell-like inequality \cite{dada2011experimental}. Tests of Bell inequalities are essential for showing a device-independent violation of local realism in high dimensions \cite{vaziri2002experimental, cai2016new}, and could be useful for performing high-dimensional quantum cryptography without the need for trusted devices.

If one is only interested in the entanglement dimensionality, the strong bounds posed by Bell-like inequalities on the required visibilities and the number of measurements can be significantly relaxed (Figure 2c). The theoretical tools for doing so are called entanglement dimensionality (or Schmidt number) witnesses \cite{terhal2000schmidt, sanpera2001schmidt}, which are generalizations of entanglement witnesses. In general, an entanglement witness is a function involving measurements performed on a given state, and is bounded by a value maximally reachable by a separable state. When the measurement outcomes exceed this bound, the state must have been non-separable and thus entangled. The premise of this is that the results of certain measurements for separable states (such as visibilities in different bases) are upper-bounded. If one exceeds this bound, the state was entangled \cite{guhne2009entanglement}. This has been applied to many two-dimensional OAM subspaces \cite{agnew2012observation}. Similarly, an entanglement dimensionality witness has bounds for a specific number of entangled dimensions. For example, an arbitrary two-dimensionally entangled state can only reach a certain maximum value of the witness for a set of measurement results. If the witness measurement exceeds this value, the state was at least three-dimensionally entangled. Such entanglement witnesses have been used to show up to 103-dimensional entanglement in a 168 dimensional two-photon system \cite{krenn2014generation}. A particularly convenient version of a dimensional witness is one that is based on the fidelity of the quantum state. It is fully state-independent, can be generalized to multi-photon experiments (see below), and has recently been shown to be highly applicable to high-dimensional entanglement certification \cite{fickler2014interface, erhard2016quantum}.

While the dimensionality of entanglement is an interesting fundamental property of the quantum state, it does not necessarily correspond to the usefulness of the state in quantum protocols. Consider the state
                                                  \begin{align}
\ket{\psi_{(2+\epsilon)d}}=\frac{\ket{0,0}+\ket{1,1}+\epsilon\ket{2,2}+\epsilon\ket{3,3}}{\sqrt{2+2\epsilon^2}},
\label{2epsilondimentanglemenet}
\end{align} 
with small $\epsilon$. Such a state is 4-dimensionally entangled, but the two photons barely share more than 1 bit of nonlocal information. The useful information is characterized by the ebits, or entangled bits, and is formally named entanglement-of-formation \cite{wootters1998entanglement}. The distribution of modes in an OAM-entangled state can be optimized to increase this quantity, and several experiments have analyzed \cite{pires2010measurement, salakhutdinov2012full} and demonstrated the shaping of the two-photon OAM spectrum \cite{romero2012increasing}. Promising theoretical investigations of SPDC crystals show that the spectrum could be further increased significantly by applying chirped phase-matching \cite{svozilik2012high}. In many cases, specific quantum states are required which need to be carefully engineered. One method to engineer specific antisymmetric high-dimensional entangled quantum states exploits Hong-Ou-Mandel interference for the precise quantum state filter \cite{zhang2016engineering}.

\subsection*{High-OAM-, Hybrid- and Hyper-entanglement}
The existence of quantum phenomena in large systems leads to apparent paradoxes first formulated in the form of Schr\"odinger's cat. While generally accepted definitions of macroscopic quantum superpositions and macroscopic entanglement are still missing, there has been much progress in creating quantum systems of larger and larger sizes \cite{leggett2002testing, arndt2014testing}, for example systems involving large masses or large spatial separation. A large difference in physical quantities or in quantum numbers has also been discussed as one potential route for creating macroscopic quantum superpositions \cite{leggett2002testing}. The OAM degree of freedom offers the possibility for creating entanglement of arbitrarily large quanta in principle. This motivates the creation of two-photon states with very \textit{high-OAM entanglement} \cite{fickler2012quantum}. The experimental idea is to create polarisation-entangled photons and transfer their polarisation information to OAM in an interferometric way. The OAM quantum number is then only limited by the quality of the hologram used. With SLMs, it was possible to create a two-dimensionally entangled state with an OAM quantum number difference of 600$\hbar$. In a follow-up experiment that replaced the SLM with spiral phase mirrors made of aluminium \cite{campbell2012generation, shen2013generation} it was possible to reach a quantum number difference of 10,000$\hbar$ \cite{fickler2016quantum}. The coupling between polarisation and OAM is not only possible in an interferometric scheme, but also with light-matter interaction in an anisotropic inhomogeneous medium. Such a device is called a q-plate \cite{marrucci2006optical, nagali2009quantum} and has been used in several quantum experiments, such as a quantum random walk with up to five steps \cite{cardano2015quantum}, non-contextuality tests with four-dimensional entangled quantum states \cite{d2013experimental}, and quantum cloning \cite{nagali2010experimental, nagali2009optimal}. 

Quantum states where the entanglement is distributed between different degrees of freedom are known as \textit{Hybrid-entangled} states. An example is the following state

\begin{align}
\ket{\psi_{\textnormal{hyb}}}=&\frac{1}{\sqrt{2}}\left(\ket{L}_A\ket{2}_B+\ket{R}_A\ket{-2}_B\right)\otimes\left(\ket{0}_A\ket{H}_B\right)\nonumber\\
=&\frac{1}{\sqrt{2}}\left(\ket{L,0}_A\ket{H,2}_B+\ket{R,0}_A\ket{H,-2}_B\right)
\label{hybridexample}
\end{align}

where A and B stand for the first and second photon, R, L, H and V stand for the polarisation, and $\pm$2 and 0 stand for the OAM of the photon. While the state is only two-dimensionally entangled, the combination of two degrees of freedom allows for a large variety of complex entangled photon states. Such an entangled state was created using a q-plate, where the polarization of one photon was entangled with the OAM of the second photon \cite{karimi2010spin}. Similarly, an interferometric method was used to create a quantum state with the polarization of one photon entangled with a complex polarization pattern (based on a combination of polarization and spatial modes) of the second photon \cite{fickler2014quantum}. These studies show the vast possibilities for quantum correlations encoded in complex ways and spread over several degrees of freedom. Note that while one can formally write single photons (and even classical Maxwell fields) with non-separable polarisation-spatial wave functions, such systems cannot be considered entangled as they are missing the crucial property of quantum entanglement: spatial separation or nonlocality (see \cite{karimi2015classical} for an illuminating discussion). Modern ICCD cameras have the possibility to detect single photons, and can be triggered in the nanosecond regime. This allows their application in quantum experiments, and in particular in experiments involving the spatial modes of photons. Using hybrid entanglement, ICCD cameras have been used to image quantum entanglement in real-time \cite{fickler2013real}, as well as remote-state preparation with OAM-polarisation hybrid states \cite{erhard2015real}.

An interesting, complementary approach is \textit{hyper-entanglement}. Such states are simultaneously entangled in their spatial mode structure, as well as in other degrees of freedom. The first demonstration of such a state was one entangled in polarisation, time, and OAM \cite{barreiro2005generation}, which is created as such directly in the SPDC process. The state can be written as   

\begin{flalign}
\ket{\psi_{\textnormal{hyper}}}= &\big(\ket{H,H} + \ket{V,V} \big)_{Pol} \otimes&\nonumber\\
&\big(\ket{-1,1} + \ket{0,0} + \ket{1,-1} \big)_{OAM} \otimes&\nonumber\\
&\big(\ket{s,s}+\ket{l,l} \big)_{Time}&
\label{hyperentanglement}
\end{flalign} 

and contains 12-dimensional entanglement distributed over three degrees of freedom. Due to the complex encoding of the entanglement, such states can be used for improved versions of quantum protocols. For example, in superdense-coding, it allows one to go beyond the conventional linear-optics channel capacity limit and reach optimal performance \cite{barreiro2008beating}.  Furthermore, hyper-entanglement has been used to implement resource-optimized versions of remote-state-preparation protocols with linear optics only \cite{graham2015superdense}.
  
\begin{figure*}[ht!]
\includegraphics[width=0.745 \textwidth]{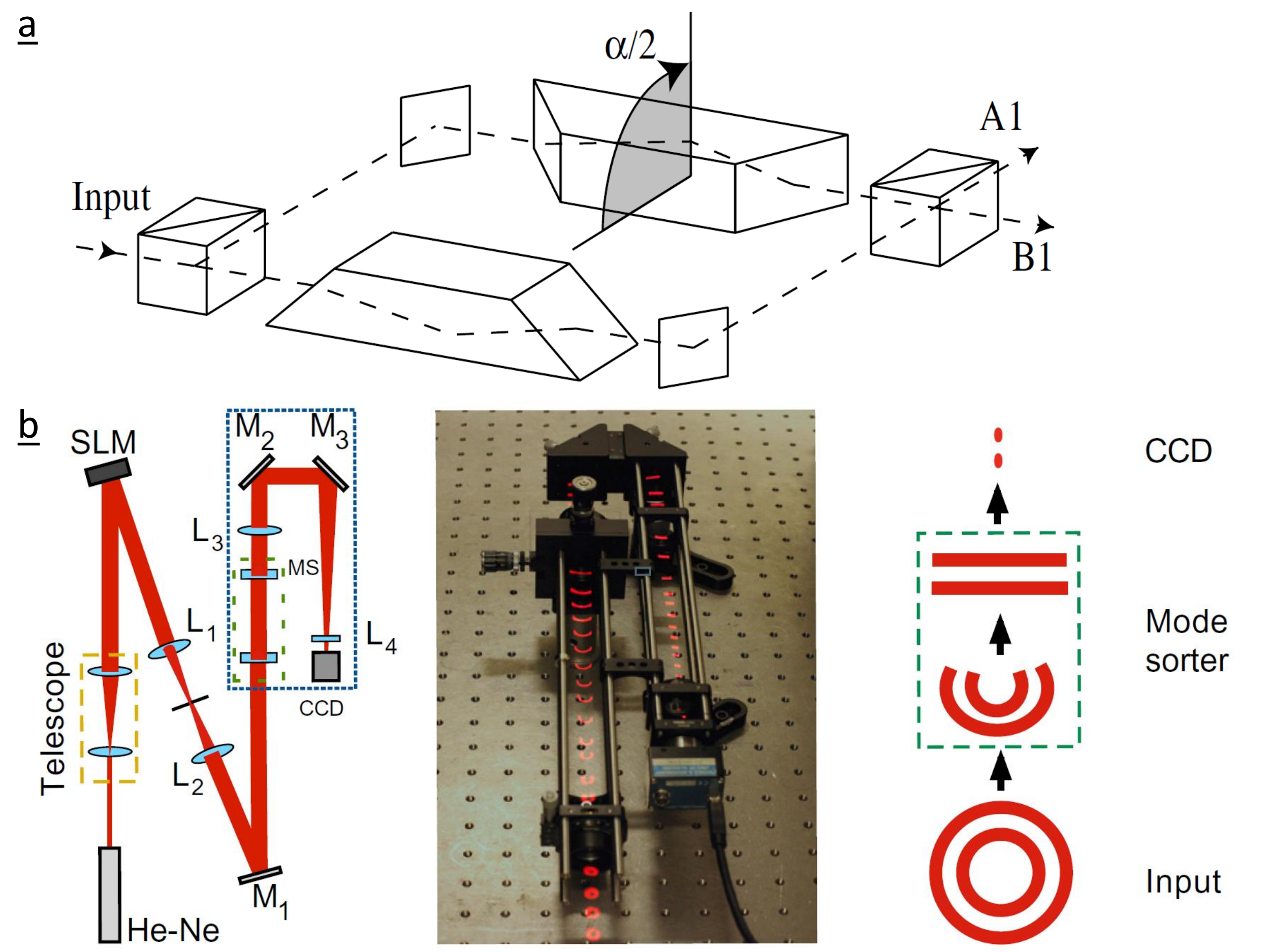}\\
\caption{The ability to measure the OAM of a single photon is very important for quantum experiments. \textbf{a} shows the idea of a non-destructive measurement of the parity of OAM in an interferometric way introduced in \cite{leach2002measuring}. Due to an OAM-dependent phase introduced by a Dove prism, even and odd modes exit from different output ports. This process can be cascaded for measuring arbitrarily large OAM spectra. (Image from \cite{leach2002measuring}) \textbf{b} shows a refractive method for efficiently separating the OAM eigenstates of a single photon. A log-polar transformation is used to convert the helical phase profile of an OAM mode into a linear phase ramp with an OAM-dependent tilt. Such modes can then be spatially separated via a simple Fourier-transforming lens. (Image from \cite{lavery2013efficient})}  
\label{fig:fig3SinglePhotons}
\end{figure*}

\subsection*{Radial Modes in quantum experiments}
Radial modes are the second quantum number of Laguerre-Gaussian modes and correspond to a radial-momentum-like property of the photon. While they haven't drawn as much attention as the OAM quantum number, they have seen an increased interest in recent years \cite{karimi2012radial, karimi2014radial, plick2015physical}. For example, several experiments have demonstrated the control of these modes and quantum correlations between them \cite{salakhutdinov2012full, krenn2014generation, geelen2013walsh, zhang2014radial}. In particular, a recent experiment demonstrated Hong-Ou-Mandel interference between two photons carrying radial modes, clearly showing their quantum behaviour \cite{karimi2014exploring}. In this experiment, single photons were impressed with radial modes up to the 9th order and interfered at a beam-splitter.

\subsection*{Measuring OAM of single photons}

The measurement technique used in most entanglement experiment using holograms or SLMs allowed one to determine the OAM content of a single photon. However, this method only allowed one to ask the following question: \textit{Is the photon in a particular OAM mode (superposition)?} In order to fully utilize the high-dimensional potential of the OAM degree of freedom, one would instead like to ask the following question: \textit{What OAM quanta is the photon carrying?} This would correspond to a measurement scheme that can distinguish among arbitrarily many OAM levels of a single photon with unit efficiency. This important question was first addressed in a 2002 experiment that used a modified Mach-Zehnder interferometer to separate even and odd quanta of OAM (Figure 3a) \cite{leach2002measuring}. Each arm of the interferometer contained a Dove prism that introduced a phase that depended both on the OAM value of the photon and the rotation angle of the prism. In this manner, a relative angle of $\alpha=\pi$ between the two Dove prisms resulted in constructive (destructive) interference for photons carrying even (odd) OAM quanta. At a different relative angle $\alpha=\frac{\pi}{2}$, the interferometer sorted between even and odd OAM quanta in multiples of two. By cascading many such interferometer elements in a clever way, one was able to efficiently measure the entire spectrum of OAM modes carried by a single photon. A recent experiment used this device to split a high-dimensional two-photon entangled state into two lower-dimensional entangled states with opposite OAM parity, demonstrating a two-particle high-dimensional analogue of the Stern-Gerlach effect \cite{erhard2016quantum}. The device can also be used as a two-input, two-output device, in a manner analogous to a polarizing beam splitter that makes it particularly useful in quantum experiments. In that scenario, the Mach-Zehnder interferometer acts as a high-dimensional OAM-parity \textit{beam splitter}, reflecting (transmitting) photons with odd (even) parity. That capability was significant for its utility in performing cyclic operations with OAM \cite{schlederer2016cyclic}, as well as in multi-photon entanglement experiments with OAM \cite{malik2016multi}, which is discussed later.

In order to sort a large number of OAM quanta, the cascaded interferometer approach is technically demanding. In 2010, a significant experiment demonstrated a two-element refractive device that separates the OAM content of a photon into its components (Figure 3b) \cite{berkhout2010efficient}. The fundamental idea consists of using a transformation that unwraps the helical phase-structure of an OAM mode into a plane wave with an OAM-dependent tilt. Thus, an $\ell=2$ mode would have twice the tilt of an $\ell=1$ mode, and so on. These tilted plane waves act similar to gratings and separate different OAM modes during propagation. In this manner, a superposition of many OAM modes can be transformed into spatially separated spots and detected individually. While the first version of this OAM sorter was realized on SLMs, a refractive version was subsequently developed that was used to sort up to 51 OAM modes \cite{lavery2012refractive, lavery2013efficient}.

The separation efficiency was later improved from 77.4\% to a theoretical value of 97.3\% by the addition of two diffractive elements that coherently copied the transformed plane-wave modes, thus reducing their overlap with neighboring ones \cite{mirhosseini2013efficient,malik2014direct}. The improved OAM sorter also allowed the separation of modes in a basis mutually unbiased with respect to the OAM basis (also known as the angular basis). This enables the measurements in two mutual unbiased bases, thus its application in various quantum applications such as high-dimensional quantum key distribution \cite{ mirhosseini2015high}.

In the first entanglement experiment involving the OAM mode sorter, this device was used in reverse to convert a high-dimensionally entangled two-photon state encoded in the path degree of freedom to one entangled in the OAM degree of freedom \cite{fickler2014interface}. It not only showed the reversibility of the sorter at the single-photon level, but also demonstrated its ability to act as a high-dimensional interface between the path and OAM degrees of freedom. The motivation is that in the path degree of freedom, it is known how to perform arbitrary unitary transformations \cite{reck1994experimental, schaeff2015experimental, carolan2015universal}. The first experiment exploiting this feature has shown the multiplication of the OAM value by an arbitrary constant integer \cite{potovcek2015quantum}. It is unknown whether such a transformation can be realized without leaving the OAM space.

\subsection*{Quantum Cryptography with Twisted Photons}
\begin{figure*}[ht!]
\includegraphics[width=0.7 \textwidth]{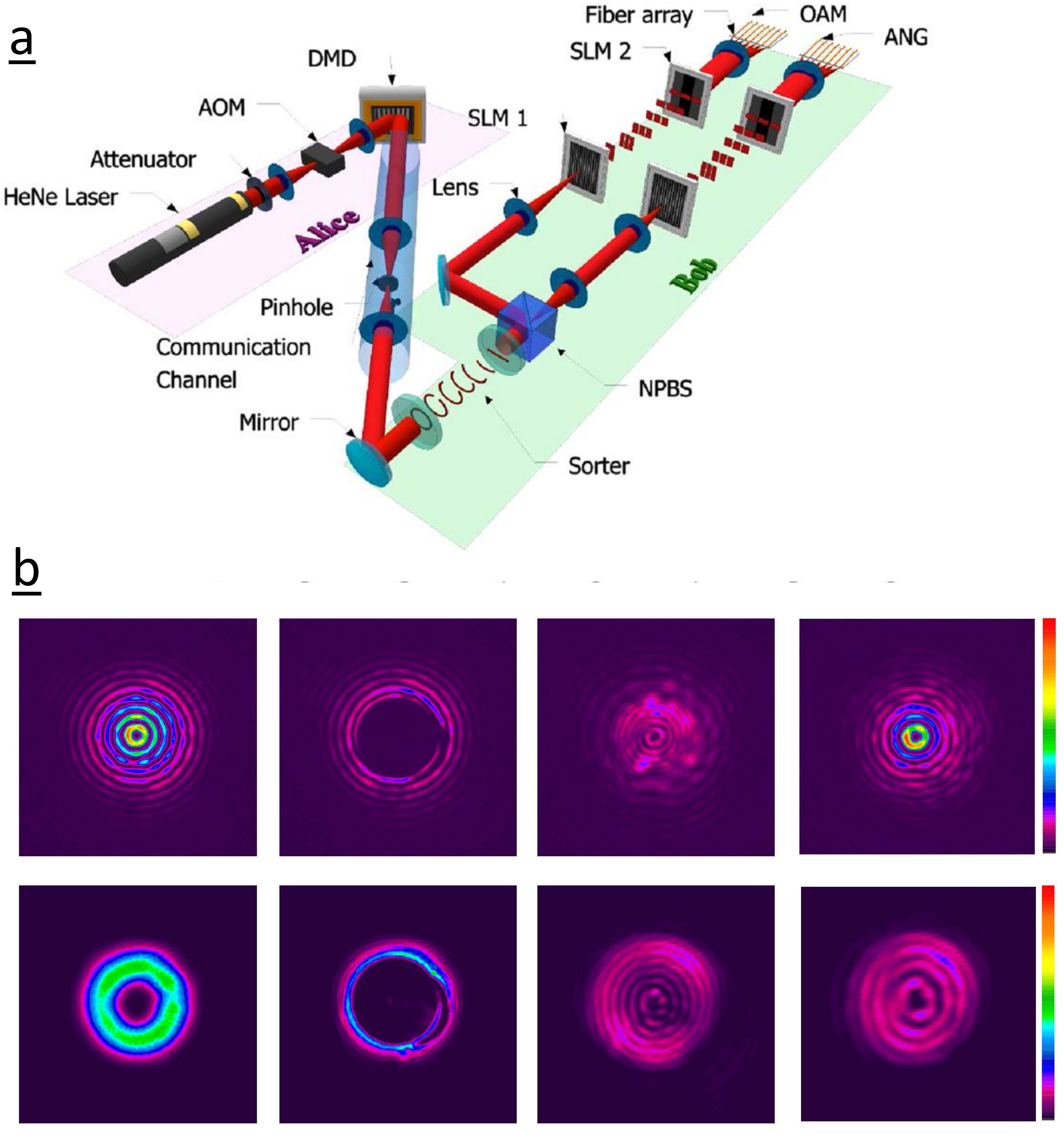}\\
\caption{\textbf{a}: A high-dimensional BB84 QKD experiment has been implemented in \cite{mirhosseini2015high}. It uses a digital micro-mirror device (DMD) for very fast encoding of spatial modes, and multi-outcome measurements in two mutual unbiased bases (OAM and angular modes). (Image from \cite{mirhosseini2015high}). \textbf{b} shows the self-healing character of Bessel-modes, which might be useful for long-distance entanglement experiments and QKD. The upper row shows Bessel modes (with nonzero OAM), and the lower one Laguerre-Gauss modes. The second and third image shows the beam after an obstruction and after 2cm of propagation. Interestingly, in the last image after 5cm of propagation, the Bessel mode reappears while the Laguerre-Gauss mode has not resumed its original structure. The self-healing property has been demonstrated for entanglement in \cite{mclaren2014self} (Image from \cite{mclaren2014self}).}  
\label{fig:fig4QKD}
\end{figure*}

Photons carrying orbital angular momentum have a natural application as carriers of quantum information, as they live in a discrete and theoretically unbounded state space. There are two primary advantages of using such a high-dimensional encoding scheme in quantum communication. First, the large state space offered by OAM allows one to send a vastly increased amount of information per photon as compared to other encoding schemes such as polarisation. The second, slightly more subtle advantage is found in an increased resistance to errors in quantum key distribution \cite{huber2013weak}. The larger the dimension of the state space, the greater the probability that an eavesdropper will introduce errors in such a communication protocol \cite{cerf2002security}. Security against individual eavesdropping attacks is also increased slightly when using the full set of mutually unbiased bases (MUBs) available in a high-dimensional space \cite{wootters1989optimal}, albeit at a reduced key rate. A recent study developed security proofs against coherent attacks on such protocols, taking into account finite-key effects \cite{sheridan2010security}.

Similar to polarisation-based protocols, OAM-based QKD can be carried out as a prepare-and-measure \cite{bennett1984quantum} scheme, or an entanglement-based \cite{ekert1991quantum} scheme. The first proof-of-principle experiment demonstrating quantum key distribution with OAM was performed with entangled qutrits in 2006 \cite{groblacher2006experimental}. Entangled photons were generated in a type-I BBO crystal and sent to two probabilistic mode analysers, which consisted of beam splitters, mode-selection holograms, and single-mode fibres. A trinary key was generated by looking for coincident detections between the two analysers, while security was demonstrated by violating a three-dimensional Bell-like inequality \cite{collins2002bell}. A recent experiment used spatial light modulators to extend this protocol up to d=5, while also using the full set of d+1 MUBs \cite{mafu2013higher}. A key component of any practical QKD system is the ability to perform multi-outcome measurements, such as those performed by a polarising beam-splitter. The development of the OAM mode sorter \cite{lavery2013efficient, mirhosseini2013efficient} opened the door to such protocols, leading to a demonstration of an OAM-based BB84 protocol in 2015 (Figure 4a) \cite{mirhosseini2015high}. In addition to multi-outcome measurements, the experiment used a digital micro-mirror device (DMD) to encode OAM modes at a rate of 4 kHz \cite{mirhosseini2013rapid}, which is two orders of magnitude higher than that possible with SLMs. An OAM-entanglement-based protocol with multi-outcome measurements is yet to be performed.

\begin{figure*}[ht!]
\includegraphics[width=0.88 \textwidth]{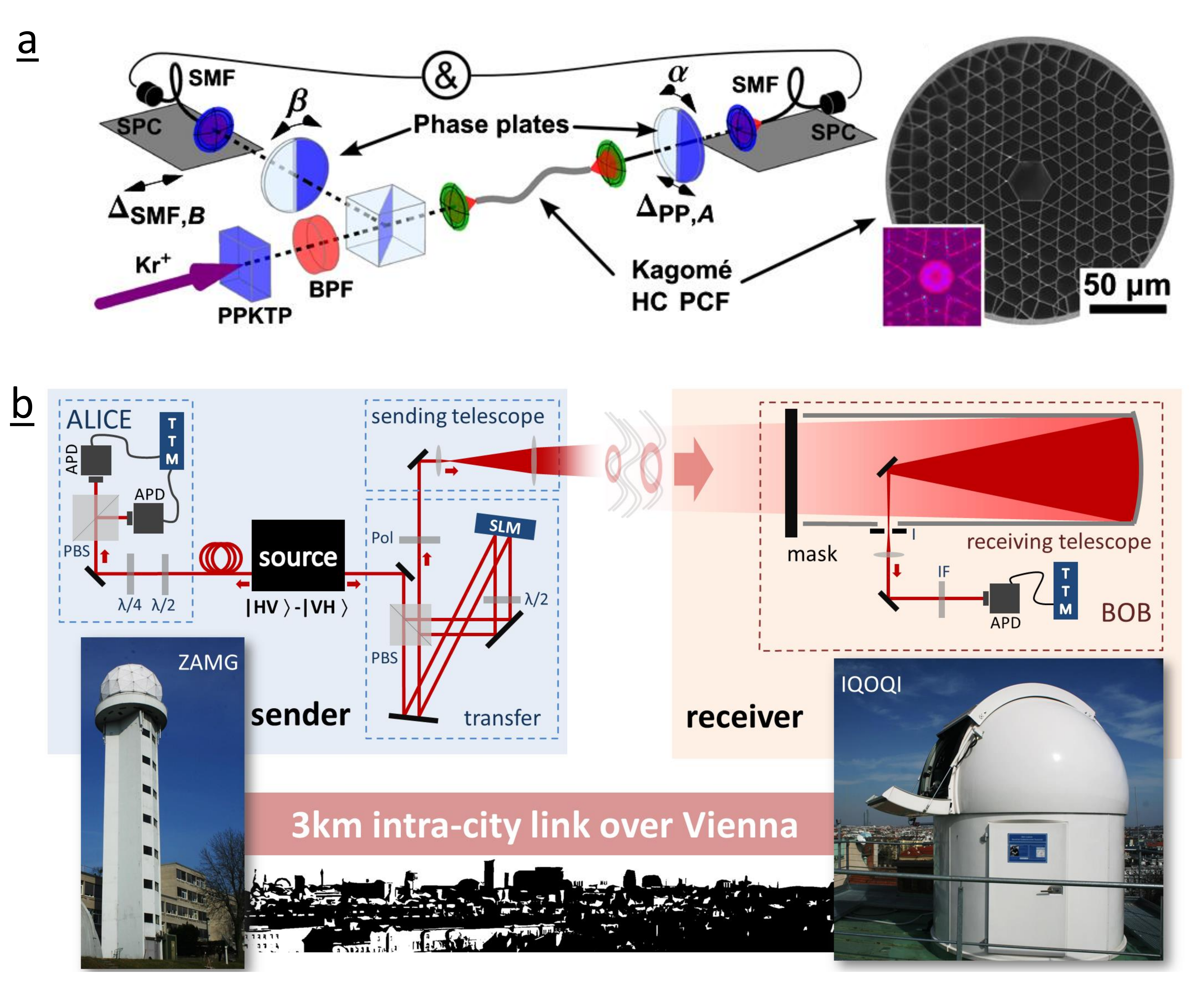}\\
\caption{Long-distance quantum communication can be done in two different ways. a shows an experiment which distributes OAM entanglement via a photonic crystal fiber \cite{PhysRevLett.106.240505}, with subsequent measurement of a Bell inequality using sector plates. While the fiber was only 30 cm long, the experiment clearly shows that entanglement can in principle be coupled into and transported via fibers. (Image from \cite{PhysRevLett.106.240505}). An alternative method is the free-space transmission of OAM modes. A 3 kilometer turbulent intra-city link has been shown to support the distribution of entanglement encoded in the first two higher-order modes \cite{krenn2015twisted}. In this experiment, a polarisation-entangled pair of photons was created, where one of the photons was measured in polarisation. The polarisation information of the second photon was transferred to OAM and transmitted over 3 kilometers, and measured using a  infront of a telescope. (Image from \cite{krenn2015twisted})}  
\label{fig:fig5LongDistance}
\end{figure*}

\subsection*{Long-Distance Transmission of OAM}
In addition to efficient generation and detection, the transmission over large distances of photons carrying OAM is necessary for any practical OAM-based quantum communication protocol. This is also important for loophole-free demonstration of All-versus-Nothing violations of local realism \cite{yang2005all}, which require large spatial separation of the photons. One can take two approaches to transmit photonic OAM over such macroscopic distances: in free-space or through fiber. In a 2012 experiment, one photon from a pair of spatially-entangled photons was transmitted through a 30cm long hollow-core photonic crystal fiber. A Bell inequality was used to confirm two-dimensional spatial-mode entanglement after the transport (Figure 5a) \cite{PhysRevLett.106.240505}. In a 2013 experiment, three lowest order OAM modes (-1, 0, and +1) carried by a laser beam were sent through 1.1km of a specially designed fiber \cite{bozinovic2013terabit}. However, these types of fibers have not been exploited in quantum experiments yet.

Extensive research has been conducted on the topic of free-space transmission of OAM quanta and OAM entanglement through turbulence, including several theoretical \cite{brunner2013robust, leonhard2015universal, roux2014parameter} and lab-scale simulations \cite{pors2011transport, ibrahim2013orbital, farias2015resilience}. These studies have predicted that the mode quality would deteriorate rapidly in turbulent atmosphere, leading to a drastic effect on OAM entanglement. Only two experiments involving single or entangled photons carrying OAM have been performed over large distances. The first experiment transmitted weak coherent laser pulses over a distance of 210 meters, and used them to perform a two-dimensional BB84 quantum cryptography protocol \cite{vallone2014free}. That experiment was performed in a large hall in Padua in order to minimize the detrimental effect of atmospheric turbulence. A second experiment transmitted two-dimensional OAM-entanglement over an outdoor intra-city environment of 3 kilometers in Vienna (Figure 5b) \cite{krenn2015twisted}. Because of sunlight and the necessity to identify individual photons, the experiment was performed during the night. An interesting method which could improve long-distance distributions of OAM entanglement is the application of self-healing Bessel-beams. Entanglement directly after an obstruction has been shown to be destroyed, however after some propagation distance, it is self-healed and can be detected again (Figure 4b) \cite{mclaren2014self}.

\begin{figure*}[ht!]
\includegraphics[width=0.88 \textwidth]{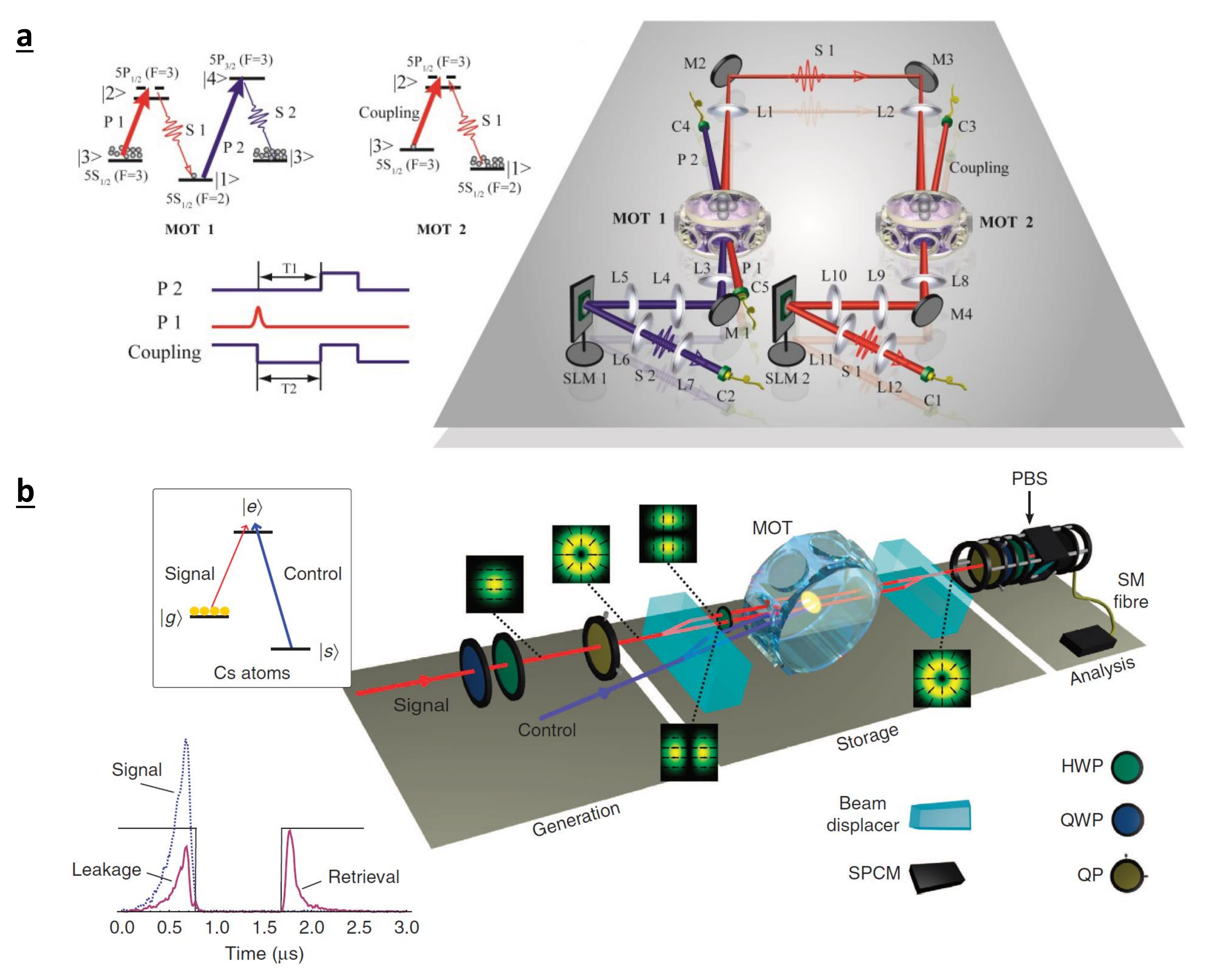}\\
\caption{\underline{Quantum memories for spatial modes.} \textbf{a}:  A quantum memory for high-dimensional entangled states spatially separated by 1 meter \cite{ding2016high}. On the left side one sees the energy-level diagram of $^{85}$Rubidium, as well as the time sequence for creating and storing entanglement in it. On the right is a sketch of the experimental setup with two distant magneto-optical traps (MOTs) between which entanglement is generated (Image from \cite{ding2016high}). \textbf{b}: A quantum memory which can store both the polarisation as well as the spatial mode information of photons is shown. The photons are stored in a cloud of Caesium atoms, and the storage time is roughly 1 $\mu$s. (Image from \cite{parigi2015storage})}  
\label{fig:fig6Atom}
\end{figure*}

\subsection*{Manipulating atoms with OAM}

The prospect of coupling the OAM carried by single photons or pairs of OAM-entangled photons to quantum states of matter remains a tantalising one. The first step in this direction was taken in 2006, when a laser beam carrying OAM was used to generate atomic vortex states in a Sodium Bose-Einstein condensate (BEC) via the process of stimulated Raman scattering (SRS) \cite{andersen2006quantized}. Interference between the different resulting vortex states demonstrated the coherent superposition of OAM modes in the BEC. In the same year, an experiment demonstrated the creation of two-dimensional OAM-entanglement between an ensemble of cold Rubidium-87 atoms and a single photon \cite{inoue2006entanglement}. A "write" laser pulse excited an atomic transition in the cloud, emitting an anti-Stokes photon entangled with the atom cloud. After a storage time of 100ns, a "read" laser pulse mapped the atomic transition back into a Stokes photon, which was shown to be OAM-entangled with the anti-Stokes photon. This constituted the first demonstration of a read-only memory for OAM based on the SRS process, and was extended to three-dimensional atom-photon entanglement a few years later \cite{inoue2009measuring}. More recently, the same scheme was expanded upon to demonstrate the storage of high-dimensional OAM-entanglement between two spatially separated atomic clouds (Figure 6a) \cite{ding2015quantum, ding2016high}. An alternative method involves the storage of high-dimensional quantum entanglement in a rare-earth crystal \cite{zhou2015quantum}. A three-dimensional Bell-inequality was violated after a storage time of 40ns, with OAM values of up to $\ell$=25. Another recent experiment demonstrated the writing, storage, and read-out of single photons and weak coherent pulses carrying OAM modes via electromagnetically-induced transparency in a cloud of cold Rubidium with a storage time of ~400ns \cite{ding2013single} and Caesium atoms with a storage time of ~1$\mu$s \cite{nicolas2014quantum}. The latter experiment has been extended in order to store both the polarisation and OAM information of the photon (Figure 6b) \cite{parigi2015storage}. The photons were retrieved with a high fidelity, demonstrating this technique's potential in quantum information schemes. In addition to manipulating atomic excitations, there have been a series of theoretical proposals for exciting quadrupole transitions in ions with light carrying OAM \cite{schmiegelow2012light, lembessis2013enhanced}, as well one recent experimental demonstration \cite{schmiegelow2015excitation}.

\subsection*{Multi-photon experiments with OAM}
\begin{figure*}[ht!]
\includegraphics[width=0.68 \textwidth]{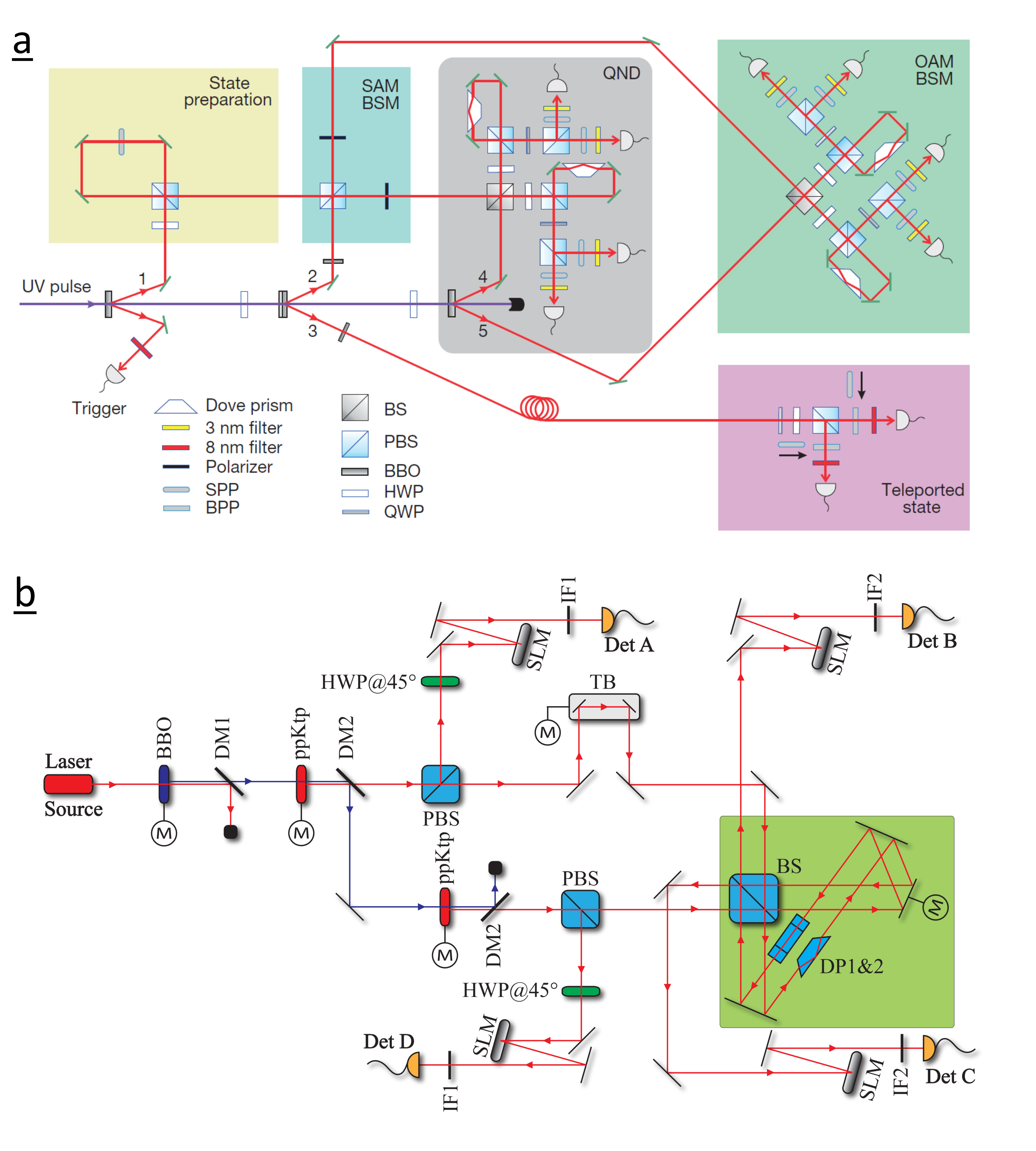}\\
\caption{Multiphoton quantum experiments involving OAM. In \textbf{a}, an experiment is shown where two properties of a photon (the polarisation and the parity of an $\ell=\pm1$ spatial mode) are teleported simultaneously. It is possible by a quantum non-demolition measurement, which curiously is implemented itself as a quantum teleportation scheme. For this, six photons are required to teleport the (2x2)-dimensional quantum state. (Image taken from \cite{wang2015quantum}). In \textbf{b} an experiment is shown which creates a genuine multipartite high-dimensional entangled state. Similarly to multiphoton polarisation experiments, the which-crystal information is erased by an interferometer that sorts even and odd OAM modes. The resulting state has an asymmetric entanglement structure -- a feature that can only exist when both the number of particles and the number of dimensions is larger than 2. (Image taken from \cite{malik2016multi}).}  
\label{fig:fig7MultiPhoton}
\end{figure*}

Over the last two years, experiments with OAM-entanglement have ventured into the challenging multi-photon regime. The first such experiment teleported a photon in a hybrid four-dimensional OAM-polarisation state space using 3 entangled photon pairs (Figure 7a) \cite{wang2015quantum}. In order to do so, the experiment implemented a unique feature -- a quantum non-demolition (QND) measurement that heralded the presence of a photon without destroying it. In order to teleport a qubit, one needs to perform a so-called Bell-state measurement (BSM) on it, which involves projecting it into an entangled state with another qubit. In that experiment, two consecutive BSMs in the polarisation and OAM state spaces were cascaded. However, for the second BSM to work, it was essential that each photon from the first BSM arrived in a different path. The QND measurement ensured that a photon was indeed on its way, and interestingly was itself implemented by quantum teleportation. In this manner, the authors were able to project into two hyper-entangled Bell states with an efficiency of 1/32, allowing them to teleport an OAM-polarisation ququart with fidelities ranging from 0.57 to 0.68.


The next two experiments explored the creation of the first multi-photon entangled states of OAM. In one experiment, a nonlinear crystal was pumped with a strong laser in order to produce two OAM-entangled photon pairs. The four photons were then probabilistically split with beam splitters, which resulted in a four-photon Dicke-state entangled in two dimensions \cite{hiesmayr2016observation}. In order to create specific multi-photon entangled states such as a Greenberger-Horne-Zeilinger state, one can combine two pairs of entangled photons in such a manner that the which-crystal information for one or more photons is erased, leading to a four-photon entangled state \cite{zeilinger1997three}. A recent experiment used the interferometric OAM beam-splitter discussed above to combine OAM-entangled pairs from two different crystals in this manner (Figure 7b) \cite{malik2016multi}. This device mixed the odd and even OAM components of the input photons, projecting them into a high-dimensional multi-photon OAM-entangled state given by $\ket{\psi}=\frac{1}{\sqrt{3}}\big(\ket{0,0,0}+\ket{1,1,1}+\ket{2,2,1}\big)$. Interestingly, the state was asymmetrically entangled -- two photons were in a three-dimensional OAM space and a third lived in two dimensions. This state is just one example out of a vast family of multipartite entangled states that are only possible when both the dimension and the number of particles are greater than two \cite{huber2013structure}. Experimental methods for creating many of these states were recently found by using a computer algorithm that combined optical components and analysed the resulting state \cite{krenn2016automated}. An interesting question that remains is how such novel types of entanglement can be used in quantum applications or fundamental tests. 

\subsection*{Outlook}
Many interesting questions on the orbital angular momentum of photons are yet to be answered and several exciting avenues for future research can be identified.

One of them is whether high-dimensionally entangled states can be transmitted over large, multi-kilometer distances. This is necessary for any practical implementation of quantum communication protocols with OAM and could be useful in fundamental studies of local realism. While classical OAM beams have been transmitted over 1 kilometer in fiber \cite{bozinovic2013terabit} and over 143 kilometers in free-space \cite{krenn2016twisted}, similar distances are yet to be achieved at the quantum level.

One significant motivation to investigate photonic OAM is its potential advantage in quantum cryptography. An important step into that direction would be a full implementation of an entanglement-based quantum key distribution protocol, involving two parties separated by a large distance. These parties would share pairs of high-dimensionally entangled photons and perform multi-outcome measurements on them. A detailed analysis of QKD implemented with OAM of photons is necessary to understand how the high-dimensional alphabet can be fully exploited to finally outperform polarisation-based QKD in terms of secure key-rate.

The investigation of the advantages offered by quantum computation with higher-level alphabets is another interesting avenue. A recent theoretical study has shown how to use ternary systems to employ the Shor algorithm, both by extending the algorithm to three dimensions, and by encoding two-dimensional quantum information in a three-dimensional system \cite{bocharov2016factoring}. That study found several advantages compared to binary quantum architectures, such as more robust encoding of quantum information. Further analysis in high-dimensional quantum circuit designs, in particular the advantages and challenges compared to binary quantum circuits would be desirable. We hope that the work reviewed here provides additional motivation for investigating possible applications of the high-dimensional degree-of-freedom in quantum computing.

Another intriguing question is how the non-local information encoded in photons (such as entanglement dimensionality and ebits) can be increased and verified in state-independent ways. Research in this direction might explore the combination of high-dimensional entanglement in frequency and spatial modes to use the full potential of space- and time encoding. The investigation on whether a fundamental limit of non-locally shared information could exist would certainly be interesting. A related question is whether such a fundamental limit could exist for two-dimensionally entangled pairs with very high OAM values, and whether general relativistic effects would play a role in this scenario \cite{tamburini2011twisting}.

Quantum systems on a new level of complexity are accessible with multi-dimensional, multi-partite, multi-degree-of-freedom entangled states and could show interesting properties and possibilities not present in simpler systems. For that, both theoretical methods to quantify the complexity as well as experimentally feasible methods for generating such states will be necessary. In this respect, new types of transformations and interfaces between different degrees-of-freedom seem desirable for the implementation of various quantum protocols. 

Coupling between high-dimensional entangled photons and matter could show interesting new physical insights. For example, the transfer or manipulation of photonic OAM to single quasiparticles such as polarons or plasmons would be intriguing. For this, nanophotonics and light-matter interactions with OAM of light \cite{ren2006plasmon, tischler2014experimental} need further investigation at the quantum level.

While advances in technology, combined with important theoretical developments, have steadily pushed the limits of quantum optics research in the field of OAM, many striking directions remain to be explored.  It is expected that research over the next years will give new answers to several of the questions raised, and -- hopefully -- will ask new exciting ones on the physical and technical properties of individual photons carrying quanta of angular momentum.

\section*{Acknowlegdements}
This work was supported by the Austrian Academy of Sciences (ÖAW), by the European Research Council (SIQS Grant No. 600645 EU-FP7-ICT) and the Austrian Science Fund (FWF) with SFB F40 (FOQUS).
\bibliographystyle{unsrt}
\bibliography{refs}

\begin{thebibliography}{100}

\bibitem{allen1992orbital}
Les Allen, Marco~W Beijersbergen, RJC Spreeuw, and JP~Woerdman.
\newblock Orbital angular momentum of light and the transformation of
  laguerre-gaussian laser modes.
\newblock {\em Physical Review A}, 45(11):8185, 1992.

\bibitem{arlt1999parametric}
J~Arlt, K~Dholakia, L~Allen, and MJ~Padgett.
\newblock Parametric down-conversion for light beams possessing orbital angular
  momentum.
\newblock {\em Physical Review A}, 59(5):3950, 1999.

\bibitem{mair2001entanglement}
Alois Mair, Alipasha Vaziri, Gregor Weihs, and Anton Zeilinger.
\newblock Entanglement of the orbital angular momentum states of photons.
\newblock {\em Nature}, 412(6844):313--316, 2001.

\bibitem{mair2000nichtlokale}
Alois~Erwin Mair.
\newblock {\em Nichtlokale und singul{\"a}re Quantenzust{\"a}nde des Lichts}.
\newblock University of Vienna, 2000.

\bibitem{torres2003quantum}
JP~Torres, A~Alexandrescu, and Lluis Torner.
\newblock Quantum spiral bandwidth of entangled two-photon states.
\newblock {\em Physical Review A}, 68(5):050301, 2003.

\bibitem{vaziri2003concentration}
Alipasha Vaziri, Jian-Wei Pan, Thomas Jennewein, Gregor Weihs, and Anton
  Zeilinger.
\newblock Concentration of higher dimensional entanglement: qutrits of photon
  orbital angular momentum.
\newblock {\em Physical review letters}, 91(22):227902, 2003.

\bibitem{molina2004triggered}
G~Molina-Terriza, A~Vaziri, J~{\v{R}}eh{\'a}{\v{c}}ek, Z~Hradil, and
  A~Zeilinger.
\newblock Triggered qutrits for quantum communication protocols.
\newblock {\em Physical review letters}, 92(16):167903, 2004.

\bibitem{oemrawsingh2005experimental}
SSR Oemrawsingh, X~Ma, D~Voigt, A~Aiello, ER~Eliel, JP~Woerdman, et~al.
\newblock Experimental demonstration of fractional orbital angular momentum
  entanglement of two photons.
\newblock {\em Physical review letters}, 95(24):240501, 2005.

\bibitem{oemrawsingh2006high}
SSR Oemrawsingh, JA~de~Jong, X~Ma, A~Aiello, ER~Eliel, JP~Woerdman, et~al.
\newblock High-dimensional mode analyzers for spatial quantum entanglement.
\newblock {\em Physical Review A}, 73(3):032339, 2006.

\bibitem{pors2008shannon}
JB~Pors, SSR Oemrawsingh, A~Aiello, MP~Van~Exter, ER~Eliel, JP~Woerdman, et~al.
\newblock Shannon dimensionality of quantum channels and its application to
  photon entanglement.
\newblock {\em Physical review letters}, 101(12):120502, 2008.

\bibitem{pors2011high}
Bart-Jan Pors, Filippo Miatto, ER~Eliel, JP~Woerdman, et~al.
\newblock High-dimensional entanglement with orbital-angular-momentum states of
  light.
\newblock {\em Journal of Optics}, 13(6):064008, 2011.

\bibitem{walborn2003multimode}
SP~Walborn, AN~De~Oliveira, S~P{\'a}dua, and CH~Monken.
\newblock Multimode hong-ou-mandel interference.
\newblock {\em Physical review letters}, 90(14):143601, 2003.

\bibitem{peeters2007orbital}
WH~Peeters, EJK Verstegen, and MP~Van~Exter.
\newblock Orbital angular momentum analysis of high-dimensional entanglement.
\newblock {\em Physical Review A}, 76(4):042302, 2007.

\bibitem{yao2006observation}
Eric Yao, Sonja Franke-Arnold, Johannes Courtial, Miles~J Padgett, and
  Stephen~M Barnett.
\newblock Observation of quantum entanglement using spatial light modulators.
\newblock {\em Optics express}, 14(26):13089--13094, 2006.

\bibitem{jack2010entanglement}
B~Jack, AM~Yao, J~Leach, J~Romero, S~Franke-Arnold, DG~Ireland, SM~Barnett, and
  MJ~Padgett.
\newblock Entanglement of arbitrary superpositions of modes within
  two-dimensional orbital angular momentum state spaces.
\newblock {\em Physical Review A}, 81(4):043844, 2010.

\bibitem{leach2009violation}
J~Leach, B~Jack, J~Romero, M~Ritsch-Marte, RW~Boyd, AK~Jha, SM~Barnett,
  S~Franke-Arnold, and MJ~Padgett.
\newblock Violation of a bell inequality in two-dimensional orbital angular
  momentum state-spaces.
\newblock {\em Optics express}, 17(10):8287--8293, 2009.

\bibitem{romero2010violation}
J~Romero, J~Leach, B~Jack, SM~Barnett, MJ~Padgett, and S~Franke-Arnold.
\newblock Violation of leggett inequalities in orbital angular momentum
  subspaces.
\newblock {\em New Journal of Physics}, 12(12):123007, 2010.

\bibitem{chen2012hardy}
Lixiang Chen and Jacquiline Romero.
\newblock Hardy's nonlocality proof using twisted photons.
\newblock {\em Optics express}, 20(19):21687--21692, 2012.

\bibitem{hiesmayr2013complementarity}
Beatrix~C Hiesmayr and Wolfgang L{\"o}ffler.
\newblock Complementarity reveals bound entanglement of two twisted photons.
\newblock {\em New Journal of Physics}, 15(8):083036, 2013.

\bibitem{mclaren2012entangled}
Melanie McLaren, Megan Agnew, Jonathan Leach, Filippus~S Roux, Miles~J Padgett,
  Robert~W Boyd, and Andrew Forbes.
\newblock Entangled bessel-gaussian beams.
\newblock {\em Optics Express}, 20(21):23589--23597, 2012.

\bibitem{mclaren2013two}
Melanie McLaren, Jacquiline Romero, Miles~J Padgett, Filippus~S Roux, and
  Andrew Forbes.
\newblock Two-photon optics of bessel-gaussian modes.
\newblock {\em Physical Review A}, 88(3):033818, 2013.

\bibitem{romero2011entangled}
J~Romero, J~Leach, B~Jack, MR~Dennis, S~Franke-Arnold, SM~Barnett, and
  MJ~Padgett.
\newblock Entangled optical vortex links.
\newblock {\em Physical review letters}, 106(10):100407, 2011.

\bibitem{krenn2013entangled}
Mario Krenn, Robert Fickler, Marcus Huber, Radek Lapkiewicz, William Plick,
  Sven Ramelow, and Anton Zeilinger.
\newblock Entangled singularity patterns of photons in ince-gauss modes.
\newblock {\em Physical Review A}, 87(1):012326, 2013.

\bibitem{agnew2011tomography}
Megan Agnew, Jonathan Leach, Melanie McLaren, F~Stef Roux, and Robert~W Boyd.
\newblock Tomography of the quantum state of photons entangled in high
  dimensions.
\newblock {\em Physical Review A}, 84(6):062101, 2011.

\bibitem{vaziri2002experimental}
Alipasha Vaziri, Gregor Weihs, and Anton Zeilinger.
\newblock Experimental two-photon, three-dimensional entanglement for quantum
  communication.
\newblock {\em Physical Review Letters}, 89(24):240401, 2002.

\bibitem{krenn2014generation}
Mario Krenn, Marcus Huber, Robert Fickler, Radek Lapkiewicz, Sven Ramelow, and
  Anton Zeilinger.
\newblock Generation and confirmation of a (100$\times$ 100)-dimensional
  entangled quantum system.
\newblock {\em Proceedings of the National Academy of Sciences},
  111(17):6243--6247, 2014.

\bibitem{langford2004measuring}
Nathan~K Langford, Rohan~B Dalton, Michael~D Harvey, Jeremy~L O'Brien,
  Geoffrey~J Pryde, Alexei Gilchrist, Stephen~D Bartlett, and Andrew~G White.
\newblock Measuring entangled qutrits and their use for quantum bit commitment.
\newblock {\em Physical review letters}, 93(5):053601, 2004.

\bibitem{giovannini2013characterization}
D~Giovannini, J~Romero, Jonathan Leach, A~Dudley, A~Forbes, and Miles~J
  Padgett.
\newblock Characterization of high-dimensional entangled systems via mutually
  unbiased measurements.
\newblock {\em Physical review letters}, 110(14):143601, 2013.

\bibitem{tonolini2014reconstructing}
Francesco Tonolini, Susan Chan, Megan Agnew, Alan Lindsay, and Jonathan Leach.
\newblock Reconstructing high-dimensional two-photon entangled states via
  compressive sensing.
\newblock {\em Scientific Reports}, (4):6542, 2014.

\bibitem{kaszlikowski2000violations}
Dagomir Kaszlikowski, Piotr Gnaci{\'n}ski, Marek {\.Z}ukowski, Wieslaw
  Miklaszewski, and Anton Zeilinger.
\newblock Violations of local realism by two entangled n-dimensional systems
  are stronger than for two qubits.
\newblock {\em Physical Review Letters}, 85(21):4418, 2000.

\bibitem{collins2002bell}
Daniel Collins, Nicolas Gisin, Noah Linden, Serge Massar, and Sandu Popescu.
\newblock Bell inequalities for arbitrarily high-dimensional systems.
\newblock {\em Physical review letters}, 88(4):040404, 2002.

\bibitem{dada2011experimental}
Adetunmise~C Dada, Jonathan Leach, Gerald~S Buller, Miles~J Padgett, and Erika
  Andersson.
\newblock Experimental high-dimensional two-photon entanglement and violations
  of generalized bell inequalities.
\newblock {\em Nature Physics}, 7(9):677--680, 2011.

\bibitem{cai2016new}
Yu~Cai, Jean-Daniel Bancal, Jacquiline Romero, and Valerio Scarani.
\newblock A new device-independent dimension witness and its experimental
  implementation.
\newblock {\em Journal of Physics A: Mathematical and Theoretical},
  49(30):305301, 2016.

\bibitem{terhal2000schmidt}
Barbara~M Terhal and Pawe{\l} Horodecki.
\newblock Schmidt number for density matrices.
\newblock {\em Physical Review A}, 61(4):040301, 2000.

\bibitem{sanpera2001schmidt}
Anna Sanpera, Dagmar Bru{\ss}, and Maciej Lewenstein.
\newblock Schmidt-number witnesses and bound entanglement.
\newblock {\em Physical Review A}, 63(5):050301, 2001.

\bibitem{guhne2009entanglement}
Otfried G{\"u}hne and G{\'e}za T{\'o}th.
\newblock Entanglement detection.
\newblock {\em Physics Reports}, 474(1):1--75, 2009.

\bibitem{agnew2012observation}
M~Agnew, J~Leach, and RW~Boyd.
\newblock Observation of entanglement witnesses for orbital angular momentum
  states.
\newblock {\em The European Physical Journal D}, 66(6):1--4, 2012.

\bibitem{fickler2014interface}
Robert Fickler, Radek Lapkiewicz, Marcus Huber, Martin~PJ Lavery, Miles~J
  Padgett, and Anton Zeilinger.
\newblock Interface between path and orbital angular momentum entanglement for
  high-dimensional photonic quantum information.
\newblock {\em Nature communications}, 5, 2014.

\bibitem{erhard2016quantum}
Manuel Erhard, Mehul Malik, and Anton Zeilinger.
\newblock A quantum router for high-dimensional entanglement.
\newblock {\em arXiv preprint arXiv:1605.05947}, 2016.

\bibitem{wootters1998entanglement}
William~K Wootters.
\newblock Entanglement of formation of an arbitrary state of two qubits.
\newblock {\em Physical Review Letters}, 80(10):2245, 1998.

\bibitem{pires2010measurement}
H~Di~Lorenzo Pires, HCB Florijn, and MP~van Exter.
\newblock Measurement of the spiral spectrum of entangled two-photon states.
\newblock {\em Physical review letters}, 104(2):020505, 2010.

\bibitem{salakhutdinov2012full}
VD~Salakhutdinov, ER~Eliel, and W~L{\"o}ffler.
\newblock Full-field quantum correlations of spatially entangled photons.
\newblock {\em Physical review letters}, 108(17):173604, 2012.

\bibitem{romero2012increasing}
J~Romero, D~Giovannini, S~Franke-Arnold, SM~Barnett, and MJ~Padgett.
\newblock Increasing the dimension in high-dimensional two-photon orbital
  angular momentum entanglement.
\newblock {\em Physical Review A}, 86(1):012334, 2012.

\bibitem{svozilik2012high}
Ji{\v{r}}{\'\i} Svozil{\'\i}k, Jan Pe{\v{r}}ina~Jr, and Juan~P Torres.
\newblock High spatial entanglement via chirped quasi-phase-matched optical
  parametric down-conversion.
\newblock {\em Physical Review A}, 86(5):052318, 2012.

\bibitem{zhang2016engineering}
Yingwen Zhang, Filippus~S Roux, Thomas Konrad, Megan Agnew, Jonathan Leach, and
  Andrew Forbes.
\newblock Engineering two-photon high-dimensional states through quantum
  interference.
\newblock {\em Science advances}, 2(2):e1501165, 2016.

\bibitem{leggett2002testing}
Anthony~J Leggett.
\newblock Testing the limits of quantum mechanics: motivation, state of play,
  prospects.
\newblock {\em Journal of Physics: Condensed Matter}, 14(15):R415, 2002.

\bibitem{arndt2014testing}
Markus Arndt and Klaus Hornberger.
\newblock Testing the limits of quantum mechanical superpositions.
\newblock {\em Nature Physics}, 10(4):271--277, 2014.

\bibitem{fickler2012quantum}
Robert Fickler, Radek Lapkiewicz, William~N Plick, Mario Krenn, Christoph
  Schaeff, Sven Ramelow, and Anton Zeilinger.
\newblock Quantum entanglement of high angular momenta.
\newblock {\em Science}, 338(6107):640--643, 2012.

\bibitem{campbell2012generation}
Geoff Campbell, Boris Hage, Ben Buchler, and Ping~Koy Lam.
\newblock Generation of high-order optical vortices using directly machined
  spiral phase mirrors.
\newblock {\em Applied optics}, 51(7):873--876, 2012.

\bibitem{shen2013generation}
Yong Shen, Geoff~T Campbell, Boris Hage, Hongxin Zou, Benjamin~C Buchler, and
  Ping~Koy Lam.
\newblock Generation and interferometric analysis of high charge optical
  vortices.
\newblock {\em Journal of Optics}, 15(4):044005, 2013.

\bibitem{fickler2016quantum}
Robert Fickler, Geoff~T Campbell, Ben~C Buchler, Ping~Koy Lam, and Anton
  Zeilinger.
\newblock Quantum entanglement of angular momentum states with quantum numbers
  up to 10010.
\newblock {\em arXiv preprint arXiv:1607.00922}, 2016.

\bibitem{marrucci2006optical}
Lorenzo Marrucci, C~Manzo, and D~Paparo.
\newblock Optical spin-to-orbital angular momentum conversion in inhomogeneous
  anisotropic media.
\newblock {\em Physical review letters}, 96(16):163905, 2006.

\bibitem{nagali2009quantum}
Eleonora Nagali, Fabio Sciarrino, Francesco De~Martini, Lorenzo Marrucci, Bruno
  Piccirillo, Ebrahim Karimi, and Enrico Santamato.
\newblock Quantum information transfer from spin to orbital angular momentum of
  photons.
\newblock {\em Physical review letters}, 103(1):013601, 2009.

\bibitem{cardano2015quantum}
Filippo Cardano, Francesco Massa, Hammam Qassim, Ebrahim Karimi, Sergei
  Slussarenko, Domenico Paparo, Corrado de~Lisio, Fabio Sciarrino, Enrico
  Santamato, Robert~W Boyd, et~al.
\newblock Quantum walks and wavepacket dynamics on a lattice with twisted
  photons.
\newblock {\em Science advances}, 1(2):e1500087, 2015.

\bibitem{d2013experimental}
Vincenzo D'Ambrosio, Isabelle Herbauts, Elias Amselem, Eleonora Nagali, Mohamed
  Bourennane, Fabio Sciarrino, and Ad{\'a}n Cabello.
\newblock Experimental implementation of a kochen-specker set of quantum tests.
\newblock {\em Physical Review X}, 3(1):011012, 2013.

\bibitem{nagali2010experimental}
Eleonora Nagali, Daniele Giovannini, Lorenzo Marrucci, Sergei Slussarenko,
  Enrico Santamato, and Fabio Sciarrino.
\newblock Experimental optimal cloning of four-dimensional quantum states of
  photons.
\newblock {\em Physical review letters}, 105(7):073602, 2010.

\bibitem{nagali2009optimal}
Eleonora Nagali, Linda Sansoni, Fabio Sciarrino, Francesco De~Martini, Lorenzo
  Marrucci, Bruno Piccirillo, Ebrahim Karimi, and Enrico Santamato.
\newblock Optimal quantum cloning of orbital angular momentum photon qubits
  through hong--ou--mandel coalescence.
\newblock {\em Nature Photonics}, 3(12):720--723, 2009.

\bibitem{karimi2010spin}
Ebrahim Karimi, Jonathan Leach, Sergei Slussarenko, Bruno Piccirillo, Lorenzo
  Marrucci, Lixiang Chen, Weilong She, Sonja Franke-Arnold, Miles~J Padgett,
  and Enrico Santamato.
\newblock Spin-orbit hybrid entanglement of photons and quantum contextuality.
\newblock {\em Physical Review A}, 82(2):022115, 2010.

\bibitem{fickler2014quantum}
Robert Fickler, Radek Lapkiewicz, Sven Ramelow, and Anton Zeilinger.
\newblock Quantum entanglement of complex photon polarization patterns in
  vector beams.
\newblock {\em Physical Review A}, 89(6):060301, 2014.

\bibitem{karimi2015classical}
Ebrahim Karimi and Robert~W Boyd.
\newblock Classical entanglement?
\newblock {\em Science}, 350(6265):1172--1173, 2015.

\bibitem{fickler2013real}
Robert Fickler, Mario Krenn, Radek Lapkiewicz, Sven Ramelow, and Anton
  Zeilinger.
\newblock Real-time imaging of quantum entanglement.
\newblock {\em Scientific reports}, 3, 2013.

\bibitem{erhard2015real}
Manuel Erhard, Hammam Qassim, Harjaspreet Mand, Ebrahim Karimi, and Robert~W
  Boyd.
\newblock Real-time imaging of spin-to-orbital angular momentum hybrid remote
  state preparation.
\newblock {\em Physical Review A}, 92(2):022321, 2015.

\bibitem{barreiro2005generation}
Julio~T Barreiro, Nathan~K Langford, Nicholas~A Peters, and Paul~G Kwiat.
\newblock Generation of hyperentangled photon pairs.
\newblock {\em Physical review letters}, 95(26):260501, 2005.

\bibitem{barreiro2008beating}
Julio~T Barreiro, Tzu-Chieh Wei, and Paul~G Kwiat.
\newblock Beating the channel capacity limit for linear photonic superdense
  coding.
\newblock {\em Nature physics}, 4(4):282--286, 2008.

\bibitem{graham2015superdense}
Trent~M Graham, Herbert~J Bernstein, Tzu-Chieh Wei, Marius Junge, and Paul~G
  Kwiat.
\newblock Superdense teleportation using hyperentangled photons.
\newblock {\em Nature communications}, 6, 2015.

\bibitem{leach2002measuring}
Jonathan Leach, Miles~J Padgett, Stephen~M Barnett, Sonja Franke-Arnold, and
  Johannes Courtial.
\newblock Measuring the orbital angular momentum of a single photon.
\newblock {\em Physical review letters}, 88(25):257901, 2002.

\bibitem{lavery2013efficient}
Martin~PJ Lavery, David~J Robertson, Anna Sponselli, Johannes Courtial,
  Nicholas~K Steinhoff, Glenn~A Tyler, Alan~E Willner, and Miles~J Padgett.
\newblock Efficient measurement of an optical orbital-angular-momentum spectrum
  comprising more than 50 states.
\newblock {\em New Journal of Physics}, 15(1):013024, 2013.

\bibitem{karimi2012radial}
Ebrahim Karimi and Enrico Santamato.
\newblock Radial coherent and intelligent states of paraxial wave equation.
\newblock {\em Optics letters}, 37(13):2484--2486, 2012.

\bibitem{karimi2014radial}
E~Karimi, RW~Boyd, P~de~la Hoz, H~de~Guise, J~{\v{R}}eh{\'a}{\v{c}}ek,
  Z~Hradil, A~Aiello, G~Leuchs, and Luis~Lorenzo S{\'a}nchez-Soto.
\newblock Radial quantum number of laguerre-gauss modes.
\newblock {\em Physical review A}, 89(6):063813, 2014.

\bibitem{plick2015physical}
William~N Plick and Mario Krenn.
\newblock Physical meaning of the radial index of laguerre-gauss beams.
\newblock {\em Physical Review A}, 92(6):063841, 2015.

\bibitem{geelen2013walsh}
D~Geelen and W~L{\"o}ffler.
\newblock Walsh modes and radial quantum correlations of spatially entangled
  photons.
\newblock {\em Optics letters}, 38(20):4108--4111, 2013.

\bibitem{zhang2014radial}
Yingwen Zhang, Filippus~S Roux, Melanie McLaren, and Andrew Forbes.
\newblock Radial modal dependence of the azimuthal spectrum after parametric
  down-conversion.
\newblock {\em Physical Review A}, 89(4):043820, 2014.

\bibitem{karimi2014exploring}
Ebrahim Karimi, Daniel Giovannini, Eliot Bolduc, Nicolas Bent, Filippo~M
  Miatto, Miles~J Padgett, and Robert~W Boyd.
\newblock Exploring the quantum nature of the radial degree of freedom of a
  photon via hong-ou-mandel interference.
\newblock {\em Physical Review A}, 89(1):013829, 2014.

\bibitem{schlederer2016cyclic}
Florian Schlederer, Mario Krenn, Robert Fickler, Mehul Malik, and Anton
  Zeilinger.
\newblock Cyclic transformation of orbital angular momentum modes.
\newblock {\em New Journal of Physics}, 18(4):043019, 2016.

\bibitem{malik2016multi}
Mehul Malik, Manuel Erhard, Marcus Huber, Mario Krenn, Robert Fickler, and
  Anton Zeilinger.
\newblock Multi-photon entanglement in high dimensions.
\newblock {\em Nature Photonics}, 10(4):248--252, 2016.

\bibitem{berkhout2010efficient}
Gregorius~CG Berkhout, Martin~PJ Lavery, Johannes Courtial, Marco~W
  Beijersbergen, and Miles~J Padgett.
\newblock Efficient sorting of orbital angular momentum states of light.
\newblock {\em Physical review letters}, 105(15):153601, 2010.

\bibitem{lavery2012refractive}
Martin~PJ Lavery, David~J Robertson, Gregorius~CG Berkhout, Gordon~D Love,
  Miles~J Padgett, and Johannes Courtial.
\newblock Refractive elements for the measurement of the orbital angular
  momentum of a single photon.
\newblock {\em Optics express}, 20(3):2110--2115, 2012.

\bibitem{mirhosseini2013efficient}
Mohammad Mirhosseini, Mehul Malik, Zhimin Shi, and Robert~W Boyd.
\newblock Efficient separation of the orbital angular momentum eigenstates of
  light.
\newblock {\em Nature communications}, 4, 2013.

\bibitem{malik2014direct}
Mehul Malik, Mohammad Mirhosseini, Martin~PJ Lavery, Jonathan Leach, Miles~J
  Padgett, and Robert~W Boyd.
\newblock Direct measurement of a 27-dimensional orbital-angular-momentum state
  vector.
\newblock {\em Nature communications}, 5, 2014.

\bibitem{mirhosseini2015high}
Mohammad Mirhosseini, Omar~S Maga{\~n}a-Loaiza, Malcolm~N O'Sullivan, Brandon
  Rodenburg, Mehul Malik, Martin~PJ Lavery, Miles~J Padgett, Daniel~J Gauthier,
  and Robert~W Boyd.
\newblock High-dimensional quantum cryptography with twisted light.
\newblock {\em New Journal of Physics}, 17(3):033033, 2015.

\bibitem{reck1994experimental}
Michael Reck, Anton Zeilinger, Herbert~J Bernstein, and Philip Bertani.
\newblock Experimental realization of any discrete unitary operator.
\newblock {\em Physical Review Letters}, 73(1):58, 1994.

\bibitem{schaeff2015experimental}
Christoph Schaeff, Robert Polster, Marcus Huber, Sven Ramelow, and Anton
  Zeilinger.
\newblock Experimental access to higher-dimensional entangled quantum systems
  using integrated optics.
\newblock {\em Optica}, 2(6):523--529, 2015.

\bibitem{carolan2015universal}
Jacques Carolan, Christopher Harrold, Chris Sparrow, Enrique
  Mart{\'\i}n-L{\'o}pez, Nicholas~J Russell, Joshua~W Silverstone, Peter~J
  Shadbolt, Nobuyuki Matsuda, Manabu Oguma, Mikitaka Itoh, et~al.
\newblock Universal linear optics.
\newblock {\em Science}, 349(6249):711--716, 2015.

\bibitem{potovcek2015quantum}
V{\'a}clav Poto{\v{c}}ek, Filippo~M Miatto, Mohammad Mirhosseini, Omar~S
  Maga{\~n}a-Loaiza, Andreas~C Liapis, Daniel~KL Oi, Robert~W Boyd, and John
  Jeffers.
\newblock Quantum hilbert hotel.
\newblock {\em Physical review letters}, 115(16):160505, 2015.

\bibitem{mclaren2014self}
Melanie McLaren, Thandeka Mhlanga, Miles~J Padgett, Filippus~S Roux, and Andrew
  Forbes.
\newblock Self-healing of quantum entanglement after an obstruction.
\newblock {\em Nature communications}, 5, 2014.

\bibitem{huber2013weak}
Marcus Huber and Marcin Paw{\l}owski.
\newblock Weak randomness in device-independent quantum key distribution and
  the advantage of using high-dimensional entanglement.
\newblock {\em Physical Review A}, 88(3):032309, 2013.

\bibitem{cerf2002security}
Nicolas~J Cerf, Mohamed Bourennane, Anders Karlsson, and Nicolas Gisin.
\newblock Security of quantum key distribution using d-level systems.
\newblock {\em Physical Review Letters}, 88(12):127902, 2002.

\bibitem{wootters1989optimal}
William~K Wootters and Brian~D Fields.
\newblock Optimal state-determination by mutually unbiased measurements.
\newblock {\em Annals of Physics}, 191(2):363--381, 1989.

\bibitem{sheridan2010security}
Lana Sheridan and Valerio Scarani.
\newblock Security proof for quantum key distribution using qudit systems.
\newblock {\em Physical Review A}, 82(3):030301, 2010.

\bibitem{bennett1984quantum}
Charles~H Bennett.
\newblock Quantum cryptography: Public key distribution and coin tossing.
\newblock In {\em International Conference on Computer System and Signal
  Processing, IEEE, 1984}, pages 175--179, 1984.

\bibitem{ekert1991quantum}
Artur~K Ekert.
\newblock Quantum cryptography based on bell's theorem.
\newblock {\em Physical review letters}, 67(6):661, 1991.

\bibitem{groblacher2006experimental}
Simon Gr{\"o}blacher, Thomas Jennewein, Alipasha Vaziri, Gregor Weihs, and
  Anton Zeilinger.
\newblock Experimental quantum cryptography with qutrits.
\newblock {\em New Journal of Physics}, 8(5):75, 2006.

\bibitem{mafu2013higher}
Mhlambululi Mafu, Angela Dudley, Sandeep Goyal, Daniel Giovannini, Melanie
  McLaren, Miles~J Padgett, Thomas Konrad, Francesco Petruccione, Norbert
  L{\"u}tkenhaus, and Andrew Forbes.
\newblock Higher-dimensional orbital-angular-momentum-based quantum key
  distribution with mutually unbiased bases.
\newblock {\em Physical Review A}, 88(3):032305, 2013.

\bibitem{mirhosseini2013rapid}
Mohammad Mirhosseini, Omar~S Magana-Loaiza, Changchen Chen, Brandon Rodenburg,
  Mehul Malik, and Robert~W Boyd.
\newblock Rapid generation of light beams carrying orbital angular momentum.
\newblock {\em Optics express}, 21(25):30196--30203, 2013.

\bibitem{PhysRevLett.106.240505}
W.~L\"offler, T.~G. Euser, E.~R. Eliel, M.~Scharrer, P.~St.~J. Russell, and
  J.~P. Woerdman.
\newblock Fiber transport of spatially entangled photons.
\newblock {\em Phys. Rev. Lett.}, 106:240505, Jun 2011.

\bibitem{krenn2015twisted}
Mario Krenn, Johannes Handsteiner, Matthias Fink, Robert Fickler, and Anton
  Zeilinger.
\newblock Twisted photon entanglement through turbulent air across vienna.
\newblock {\em Proceedings of the National Academy of Sciences},
  112(46):14197--14201, 2015.

\bibitem{yang2005all}
Tao Yang, Qiang Zhang, Jun Zhang, Juan Yin, Zhi Zhao, Marek {\.Z}ukowski,
  Zeng-Bing Chen, and Jian-Wei Pan.
\newblock All-versus-nothing violation of local realism by two-photon,
  four-dimensional entanglement.
\newblock {\em Physical review letters}, 95(24):240406, 2005.

\bibitem{bozinovic2013terabit}
Nenad Bozinovic, Yang Yue, Yongxiong Ren, Moshe Tur, Poul Kristensen, Hao
  Huang, Alan~E Willner, and Siddharth Ramachandran.
\newblock Terabit-scale orbital angular momentum mode division multiplexing in
  fibers.
\newblock {\em Science}, 340(6140):1545--1548, 2013.

\bibitem{brunner2013robust}
Tobias Br{\"u}nner and Filippus~S Roux.
\newblock Robust entangled qutrit states in atmospheric turbulence.
\newblock {\em New Journal of Physics}, 15(6):063005, 2013.

\bibitem{leonhard2015universal}
Nina~D Leonhard, Vyacheslav~N Shatokhin, and Andreas Buchleitner.
\newblock Universal entanglement decay of photonic-orbital-angular-momentum
  qubit states in atmospheric turbulence.
\newblock {\em Physical Review A}, 91(1):012345, 2015.

\bibitem{roux2014parameter}
Filippus~S Roux, Thomas Konrad, et~al.
\newblock Parameter dependence in the atmospheric decoherence of modally
  entangled photon pairs.
\newblock {\em Physical Review A}, 90(5):052115, 2014.

\bibitem{pors2011transport}
Bart-Jan Pors, CH~Monken, Eric~R Eliel, and JP~Woerdman.
\newblock Transport of orbital-angular-momentum entanglement through a
  turbulent atmosphere.
\newblock {\em Optics express}, 19(7):6671--6683, 2011.

\bibitem{ibrahim2013orbital}
A~Hamadou Ibrahim, Filippus~S Roux, Melanie McLaren, Thomas Konrad, and Andrew
  Forbes.
\newblock Orbital-angular-momentum entanglement in turbulence.
\newblock {\em Physical Review A}, 88(1):012312, 2013.

\bibitem{farias2015resilience}
Osvaldo~Jim{\'e}nez Far{\'\i}as, Vincenzo D'Ambrosio, Caterina Taballione,
  Fabrizio Bisesto, Sergei Slussarenko, Leandro Aolita, Lorenzo Marrucci,
  Stephen~P Walborn, and Fabio Sciarrino.
\newblock Resilience of hybrid optical angular momentum qubits to turbulence.
\newblock {\em Scientific reports}, 5, 2015.

\bibitem{vallone2014free}
Giuseppe Vallone, Vincenzo D'Ambrosio, Anna Sponselli, Sergei Slussarenko,
  Lorenzo Marrucci, Fabio Sciarrino, and Paolo Villoresi.
\newblock Free-space quantum key distribution by rotation-invariant twisted
  photons.
\newblock {\em Physical review letters}, 113(6):060503, 2014.

\bibitem{ding2016high}
Dong-Sheng Ding, Wei Zhang, Shuai Shi, Zhi-Yuan Zhou, Yan Li, Bao-Sen Shi, and
  Guang-Can Guo.
\newblock High-dimensional entanglement between distant atomic-ensemble
  memories.
\newblock {\em Nature Light: Science \& Applications}, 2016.

\bibitem{parigi2015storage}
Valentina Parigi, Vincenzo D'Ambrosio, Christophe Arnold, Lorenzo Marrucci,
  Fabio Sciarrino, and Julien Laurat.
\newblock Storage and retrieval of vector beams of light in a
  multiple-degree-of-freedom quantum memory.
\newblock {\em Nature communications}, 6, 2015.

\bibitem{andersen2006quantized}
MF~Andersen, Changhyun Ryu, Pierre Clad{\'e}, Vasant Natarajan, A~Vaziri,
  Kristian Helmerson, and William~D Phillips.
\newblock Quantized rotation of atoms from photons with orbital angular
  momentum.
\newblock {\em Physical review letters}, 97(17):170406, 2006.

\bibitem{inoue2006entanglement}
R~Inoue, N~Kanai, T~Yonehara, Y~Miyamoto, M~Koashi, and M~Kozuma.
\newblock Entanglement of orbital angular momentum states between an ensemble
  of cold atoms and a photon.
\newblock {\em Physical Review A}, 74(5):053809, 2006.

\bibitem{inoue2009measuring}
R~Inoue, T~Yonehara, Y~Miyamoto, M~Koashi, and M~Kozuma.
\newblock Measuring qutrit-qutrit entanglement of orbital angular momentum
  states of an atomic ensemble and a photon.
\newblock {\em Physical review letters}, 103(11):110503, 2009.

\bibitem{ding2015quantum}
Dong-Sheng Ding, Wei Zhang, Zhi-Yuan Zhou, Shuai Shi, Guo-Yong Xiang, Xi-Shi
  Wang, Yun-Kun Jiang, Bao-Sen Shi, and Guang-Can Guo.
\newblock Quantum storage of orbital angular momentum entanglement in an atomic
  ensemble.
\newblock {\em Physical review letters}, 114(5):050502, 2015.

\bibitem{zhou2015quantum}
Zong-Quan Zhou, Yi-Lin Hua, Xiao Liu, Geng Chen, Jin-Shi Xu, Yong-Jian Han,
  Chuan-Feng Li, and Guang-Can Guo.
\newblock Quantum storage of three-dimensional orbital-angular-momentum
  entanglement in a crystal.
\newblock {\em Physical review letters}, 115(7):070502, 2015.

\bibitem{ding2013single}
Dong-Sheng Ding, Zhi-Yuan Zhou, Bao-Sen Shi, and Guang-Can Guo.
\newblock Single-photon-level quantum image memory based on cold atomic
  ensembles.
\newblock {\em Nature communications}, 4, 2013.

\bibitem{nicolas2014quantum}
A~Nicolas, L~Veissier, L~Giner, E~Giacobino, D~Maxein, and J~Laurat.
\newblock A quantum memory for orbital angular momentum photonic qubits.
\newblock {\em Nature Photonics}, 8(3):234--238, 2014.

\bibitem{schmiegelow2012light}
Christian~Tom{\'a}s Schmiegelow and Ferdinand Schmidt-Kaler.
\newblock Light with orbital angular momentum interacting with trapped ions.
\newblock {\em The European Physical Journal D}, 66(6):1--9, 2012.

\bibitem{lembessis2013enhanced}
VE~Lembessis and M~Babiker.
\newblock Enhanced quadrupole effects for atoms in optical vortices.
\newblock {\em Physical review letters}, 110(8):083002, 2013.

\bibitem{schmiegelow2015excitation}
Christian~T Schmiegelow, Jonas Schulz, Henning Kaufmann, Thomas Ruster,
  Ulrich~G Poschinger, and Ferdinand Schmidt-Kaler.
\newblock Excitation of an atomic transition with a vortex laser beam.
\newblock {\em arXiv preprint arXiv:1511.07206}, 2015.

\bibitem{wang2015quantum}
Xi-Lin Wang, Xin-Dong Cai, Zu-En Su, Ming-Cheng Chen, Dian Wu, Li~Li, Nai-Le
  Liu, Chao-Yang Lu, and Jian-Wei Pan.
\newblock Quantum teleportation of multiple degrees of freedom of a single
  photon.
\newblock {\em Nature}, 518(7540):516--519, 2015.

\bibitem{hiesmayr2016observation}
BC~Hiesmayr, MJA de~Dood, and W~L{\"o}ffler.
\newblock Observation of four-photon orbital angular momentum entanglement.
\newblock {\em Physical review letters}, 116(7):073601, 2016.

\bibitem{zeilinger1997three}
Anton Zeilinger, Michael~A Horne, Harald Weinfurter, and Marek {\.Z}ukowski.
\newblock Three-particle entanglements from two entangled pairs.
\newblock {\em Physical review letters}, 78(16):3031, 1997.

\bibitem{huber2013structure}
Marcus Huber and Julio~I de~Vicente.
\newblock Structure of multidimensional entanglement in multipartite systems.
\newblock {\em Physical review letters}, 110(3):030501, 2013.

\bibitem{krenn2016automated}
Mario Krenn, Mehul Malik, Robert Fickler, Radek Lapkiewicz, and Anton
  Zeilinger.
\newblock Automated search for new quantum experiments.
\newblock {\em Physical review letters}, 116(9):090405, 2016.

\bibitem{krenn2016twisted}
Mario Krenn, Johannes Handsteiner, Matthias Fink, Robert Fickler, Rupert Ursin,
  Mehul Malik, and Anton Zeilinger.
\newblock Twisted light transmission over 143 kilometers.
\newblock {\em arXiv preprint arXiv:1606.01811}, 2016.

\bibitem{bocharov2016factoring}
Alex Bocharov, Martin Roetteler, and Krysta~M Svore.
\newblock Factoring with qutrits: Shor's algorithm on ternary and metaplectic
  quantum architectures.
\newblock {\em arXiv preprint arXiv:1605.02756}, 2016.

\bibitem{tamburini2011twisting}
Fabrizio Tamburini, Bo~Thid{\'e}, Gabriel Molina-Terriza, and Gabriele Anzolin.
\newblock Twisting of light around rotating black holes.
\newblock {\em Nature Physics}, 7(3):195--197, 2011.

\bibitem{ren2006plasmon}
Xi-Feng Ren, Guo-Ping Guo, Yun-Feng Huang, Chuan-Feng Li, and Guang-Can Guo.
\newblock Plasmon-assisted transmission of high-dimensional orbital
  angular-momentum entangled state.
\newblock {\em EPL (Europhysics Letters)}, 76(5):753, 2006.

\bibitem{tischler2014experimental}
Nora Tischler, Ivan Fernandez-Corbaton, Xavier Zambrana-Puyalto, Alexander
  Minovich, Xavier Vidal, Mathieu~L Juan, and Gabriel Molina-Terriza.
\newblock Experimental control of optical helicity in nanophotonics.
\newblock {\em Light: Science and Applications}, 3(6):e183, 2014.

\end{thebibliography}
\end{document}